\documentclass[10pt, draftcls, onecolumn, journal]{IEEEtran}
\usepackage{cite, graphicx,amssymb,color, slashbox, pict2e, multirow, rotating}
\usepackage{graphicx,amssymb}
\usepackage[tbtags]{amsmath}
\usepackage{times,amsmath}
\usepackage{subfig}
\usepackage{flushend}
\usepackage{amsmath}
\usepackage{epsfig}
\usepackage{amssymb}
\usepackage{hyperref}
\usepackage{color}
\usepackage{psfrag}
\usepackage{caption}
\usepackage{subfig}
\usepackage[usenames,dvipsnames]{xcolor}
\usepackage{textcomp}

\begin{document}
\vspace{-12mm}
\title{Sum-Rate Maximization with Minimum Power Consumption for MIMO DF Two-Way Relaying: Part I - Relay Optimization}
\vspace{-5mm}
\author{\IEEEauthorblockN{Jie Gao,
Sergiy A. Vorobyov,
Hai Jiang,
Jianshu Zhang,
and Martin Haardt
\vspace{-8mm}
}
\thanks{S. A. Vorobyov is the corresponding author. J. Gao, S. A. Vorobyov, and H. Jiang are with the Department of
Electrical and Computer Engineering, University of Alberta,
Edmonton, AB, T6G 2V4 Canada; e-mails: {\tt\{jgao3, svorobyo,
hai1\}}@ualberta.ca. J. Zhang and M. Haardt are with the
Communication Research Laboratory, Ilmenau University of
Technology, Ilmenau, 98693, Germany; e-mails: {\tt\{jianshu.zhang,
martin.haardt\}}@tu-ilmenau.de.

Some preliminary results of this paper were presented at
GLOBECOM~2012, Anaheim, CA, USA.}}

\maketitle

\begin{abstract}
The problem of power allocation is studied for a multiple-input multiple-output (MIMO) decode-and-forward (DF) two-way relaying system consisting of two source nodes and one relay. It is shown that achieving maximum sum-rate in such a system does not necessarily demand the consumption of all available power at the relay. Instead, the maximum sum-rate can be
achieved through efficient power allocation with minimum power consumption. Deriving such power allocation, however, is
nontrivial due to the fact that it generally leads to a nonconvex problem. In Part I of this two-part paper, a sum-rate maximizing power allocation with minimum power consumption is found for MIMO DF two-way relaying, in which the relay optimizes its own power allocation strategy given the power allocation strategies of the source nodes. An algorithm is proposed for efficiently finding the optimal power allocation of the relay based on the proposed idea of relative water-levels. The considered scenario features low complexity due to the fact that the relay optimizes its power allocation without coordinating the source nodes. As a trade-off for the low complexity, it is shown that there can be waste of power at the source nodes because of no coordination between the relay and the source nodes. Simulation results demonstrate the performance of the proposed algorithm and the effect of asymmetry on the considered system. 
\end{abstract}

\IEEEpeerreviewmaketitle

\section{Introduction}

Two-way relaying (TWR) has recently attracted significant
interests \cite{TWR_PRT}-\cite{Asym2}. Establishing bi-directional
links between one relay and two source nodes, the information
exchange between the source nodes can be accomplished in two time
slots \cite{TWR_PRT}. In the first time slot (first phase) the
source nodes simultaneously transmit their messages to the relay
while in the second time slot (second phase) the relay forwards
the messages to the destinations. The first phase is called the
multiple access (MA) phase while the second phase is the
broadcasting (BC) phase of TWR. Compared to conventional one-way
relaying, which needs four time slots for the information exchange
between the source nodes, TWR can achieve a higher spectral
efficiency \cite{TWR_PRT}.

As the performance of TWR depends on the transmit strategies of
both the source nodes and the relay, optimizing the transmit
strategies such as power allocation and beamforming is one of the
main research interests in TWR. The transmit strategies of the relay and source
nodes depend on the relaying scheme. Similar to one-way relaying,
the relaying scheme in TWR can be amplify-and-forward (AF),
decode-and-forward (DF), etc., depending on the manner that the
received information is processed at the relay before it is
forwarded to the destinations. In the AF TWR scheme, the relay
amplifies and broadcasts the signals received from the source
nodes while it also amplifies and forwards the noise at the relay.
Sum-rate maximization for multiple-input multiple-output (MIMO) AF
TWR in which the relay and the source nodes all occupy multiple
antennas is investigated in \cite{AF2}-\cite{AF4}, while a mean
square error minimizing scheme for MIMO AF TWR is studied in
\cite{AF5}. For MIMO AF relaying, low-complexity sub-optimal
solutions can be obtained through diagonalizing the MIMO channel
based on the singular value decomposition (SVD) or the generalized SVD
(GSVD) and thereby transferring the problem of
beamforming/precoding to the problem of power allocation
\cite{AF3}, \cite{AF5}. Finding the optimal solution, on the other
hand, usually requires iterative algorithms with high complexity
\cite{AF4}, \cite{AF5}. The main challenge in investigating AF
TWR, especially AF MIMO TWR, is the strong coupling between the
transmit strategies of the source nodes and the relay due to noise
propagation. As the result of noise propagation, the optimization
over the transmit strategies of the source nodes and the relay
usually leads to nonconvex problems. For example, the information
rate of the communication in either direction is a nonconvex
function of the covariance/beamforming matrices of the sources and
the relay \cite{TWR_PRT}.

Unlike AF relaying, DF relaying does not have the problem
of noise propagation. As a result, DF TWR may achieve a better
performance than AF TWR, especially at low signal-to-noise ratios
(SNRs), at the cost of higher complexity. Moreover, optimizing the
power allocation in DF relaying usually leads to convex problems
(see for example \cite{DFow} and \cite{DF3}). DF TWR has been
studied in \cite{DF6}-\cite{DF5}. The optimal power allocation for
DF TWR is studied under a fairness constraint in \cite{DF1}. The
optimal time division between the MA and BC phases and the optimal
distribution of the relay's power for achieving weighted sum-rate
maximization are studied in \cite{DF2}. While the above two works
assume a single antenna at both the sources and the relay, the case
with multiple antennas at all nodes is investigated in
\cite{DF4}-\cite{DF5}. The achievable rate region and the optimal
transmit strategies of both the source nodes and the relay are
studied in \cite{DF4}, where the relay's optimal transmit strategy
is found by two water-filling based solutions coupled by the
relay's power limit. The authors of \cite{DF5} specifically
investigate the optimal transmit strategy in the BC phase of the
MIMO DF TWR. It is shown that there may exist different strategies
that lead to the same point in the rate region. Given that TWR can
achieve a high spectral efficiency, it is of interest to optimize
the power allocation so that the TWR scheme achieves high spectral
efficiency using minimum power consumption. Unlike AF TWR, in
which the sum-rate can always be increased when the relay has more
transmission power, the maximum sum-rate of DF TWR can be achieved
without consuming all the available power at the relay. However, finding
the sum-rate maximizing power allocation with minimum power consumption is no longer a convex
problem in general. Part I of this two-part paper studies the problem of \textcolor{Black}{finding the optimal relay power allocation which minimizes the relay power consumption among all relay power allocations that achieve the maximum sum-rate for the MIMO DF TWR given the power allocation of the source nodes. For brevity, this problem is called the sum-rate maximization with minimum (relay) power consumption.} The considered scenario is referred to as {\it relay optimization scenario}. The objective of Part I of this two-part paper is to find the optimal power allocation strategy of the relay in the relay optimization scenario.\footnote{Some preliminary results were presented at a conference \cite{Jie2012}.} The contributions of this part are as follows.

First, we show that the considered problem of sum-rate maximization with minimum relay power consumption is nonconvex. As the minimization of the relay power consumption is considered, the problem becomes more complex and the method used for deriving the optimal relay power allocation strategy in \cite{DF3} and \cite{DF4} is no longer valid. We first prove the sufficient and necessary condition for a relay power allocation to be optimal in the considered relay optimization scenario.  Then, based on this condition, we propose an efficient algorithm for finding the optimal solution. The proposed algorithm can obtain the optimal relay  power allocation in several steps without iterations, i.e., low complexity is achieved. 

Second, we show that while the relay optimization scenario has the advantage of low complexity, as a trade-off it may lead to a waste of power at the source nodes because of the lack of coordination between the source nodes and the relay. We analyze the solution of the relay optimization problem for different relay power limits and show that a waste of power at the source nodes happens when the relay has a power limit less than a certain threshold for each considered system configuration and the thresholds are also given.

Third, the effect of asymmetry on the considered MIMO DF TWR is analyzed and demonstrated. It has been observed in \cite{Asym1}, \cite{Asym2} that the asymmetry on channel gain, relay's location, etc., can cause a performance degradation in single-input single-output (SISO) TWR. We extend this to the MIMO case and show the effect of asymmetry in power limits and number of antennas at the source nodes through analysis and simulations.

The rest of the paper is organized as follows.
Section~\ref{s:sysm} gives the system model of this work. The
relay optimization problem is
solved and the features of the solution are investigated in Section~\ref{s:relayopt}. Simulation results are shown in
Section~\ref{s:simula}, and Section~\ref{s:conclu} concludes the
paper. Section~\ref{s:appen} ``Appendix'' provides proofs for some lemmas and all theorems.

\section{System Model}\label{s:sysm}
Consider a TWR with two source nodes and one relay, where source
node $i\,(i=1,2)$ and the relay have $n_i$ and $n_{\rm{r}}$
antennas, respectively. In the MA phase, source node $i$ transmits
signal $\mathbf{W}_i\mathbf{s}_i$ to the relay. Here
$\mathbf{W}_i$ is the precoding matrix of source node $i$ and
$\mathbf{s}_i$ is the complex Gaussian information symbol vector of source node $i$.
\textcolor{Black}{The elements of $\mathbf{s}_i, \forall i$ are
independent and identically distributed with zero mean and unit variance.} 
The channels from source node $i$
to the relay and from the relay to source node $i$ are denoted as
$\mathbf{H}_{i\rm{r}}$ and $\mathbf{H}_{{\rm{r}} i}$,
respectively. Receiver channel state information is assumed at both the relay and the source nodes, i.e., source node $i$ knows $\mathbf{H}_{{\rm{r}} i}$ and the relay knows $\mathbf{H}_{i\rm{r}}, \forall i$. It is also assumed that the relay knows $\mathbf{H}_{{\rm{r}} i}, \forall i$ by using either channel reciprocity or
channel feedback. The received signal at the relay in the MA
phase is
\begin{equation}\label{e:eqmac}
\mathbf{y}_{\rm r} = \mathbf{H}_{1\rm{r}} \mathbf{W}_1
\mathbf{s}_1 + \mathbf{H}_{2\rm{r}} \mathbf{W}_2 \mathbf{s}_2 +
\mathbf{n}_{\rm{r}}
\end{equation}
where $\mathbf{n}_{\rm{r}}$ is the noise at the relay with
covariance matrix $\sigma_{\rm{r}}^2\mathbf{I}$ in which
$\mathbf{I}$ denotes the identity matrix. The maximum
transmission power of source node $i$ is limited to $P_i^{\rm
max}$. Define the transmit covariance matrices $\mathbf{D}_i =
\mathbf{W}_i \mathbf{W}_i^{\rm H}, \forall i$, in which $(\cdot)^{\rm H}$ stands for the
conjugate transpose, and let $\mathbf{D}
= [\mathbf{D}_1, \mathbf{D}_2]$. Then the sum-rate of the MA phase
is bounded by \cite{Mimo_Cap}
\begin{eqnarray}\label{e:eqmacR}
\hspace{3mm}R^{\rm ma}(\mathbf{D}) 
= \log{\bigg|\mathbf{I}\!+\!(\mathbf{H}_{1{\rm{r}}}
\mathbf{D}_1 \mathbf{H}_{1{\rm{r}}}^{\rm H}\!+\!
\mathbf{H}_{2{\rm{r}}} \mathbf{D}_2 \mathbf{H}_{2{\rm{r}}}^{\rm
H}) (\sigma_{{\rm{r}}}^2)^{-1}\bigg|}
\end{eqnarray}
where $|\cdot|$ denotes determinant. In the BC phase, the relay decodes $\mathbf{s}_1$ and
$\mathbf{s}_2$ from the received signal, re-encodes messages using
superposition coding and transmits the signal
\begin{equation}\label{e:eqxr}
\mathbf{x}_{\rm{r}}=\mathbf{T}_{{\rm{r}}2}\mathbf{s}_1 +
\mathbf{T}_{{\rm{r}}1} \mathbf{s}_2
\end{equation}
where $\mathbf{T}_{{\rm{r}}i}$ is the $n_{\rm{r}}\times
n_{j}$ relay precoding matrix for relaying the signal from
source node $j$ to source node $i$.\footnote{It is assumed as
default throughout the paper that the user index $i$ and $j$
satisfy $i\neq j$.} The maximum transmission power of the relay
is limited to $P_{\rm{r}}^{\rm max}$.
\textcolor{Black}{
Note that in addition to the above superposition coding, the Exclusive-OR (XOR) based network coding is also used at the relay in the literature \cite{XORCmp1}-\cite{XORCmp3}. While XOR based network coding may achieve a better performance than superposition coding, it relies on the symmetry of the traffic from the two source nodes. The asymmetry in the traffic in the two directions can lead to a significant degradation in the performance of XOR in TWR \cite{XORCmp2}, \cite{XORCmp3}. As the general case of TWR is considered and there is no guarantee of traffic symmetry, the approach of symbol-level superposition is assumed here at the relay as it is considered in \cite{TWR_PRT} and \cite{DF2}. Moreover, for the MIMO case as considered in this work, the superposition scheme can take advantage of the MIMO channels. In the superposition scheme, the relay uses separate beamformers for the signals towards two directions, which guarantees that each transmitted signal is optimal (subject to the transmission power constraints) given its MIMO channel. This cannot be achieved if the relay uses XOR based network coding.
}

The received signal at source node $i$ is
\begin{equation}\label{e:eqbc}
\mathbf{y}_i^{\prime}=\mathbf{H}_{{\rm{r}}i}\mathbf{x}_{\rm{r}} +
\mathbf{n}_i
\end{equation}
where $\mathbf{n}_i$ is the noise at source node $i$ with
covariance matrix $\sigma_i^2\mathbf{I}$. With the knowledge of
$\mathbf{H}_{{\rm{r}}i}$ and $\mathbf{T}_{{\rm{r}}j}$, source node
$i$ subtracts the self-interference $\mathbf{H}_{{\rm{r}}i}
\mathbf{T}_{{\rm{r}}j}\mathbf{s}_i$ from the received signal and
the equivalent received signal at source node $i$ is
\begin{equation}\label{e:eqbc_Subtract}
\mathbf{y}_i = \mathbf{H}_{{\rm{r}}i} \mathbf{T}_{{\rm{r}}i}
\mathbf{s}_j + \mathbf{n}_i.
\end{equation}
Define $\mathbf{B}_i = \mathbf{T}_{{\rm{r}}i}
\mathbf{T}_{{\rm{r}}i}^{\rm H}, \forall i$ and let $\mathbf{B} = [\mathbf{B}_1, \mathbf{B}_2]$. \textcolor{Black}
{The sum-rate of the
considered DF TWR can be written as \cite{TWR_PRT}, \cite{DF2}, \cite{XORCmp1}
\begin{equation}\label{e:eqRtw}
R^{\rm tw}(\mathbf{B}, \mathbf{D}) = \frac{1}{2}\min\{R^{\rm
ma}(\mathbf{D}), R(\mathbf{B}, \mathbf{D})\}
\end{equation}
where
\begin{eqnarray}\label{e:RBD}
R (\mathbf{B},\mathbf{D})= \min\{\hat{R}_{{\rm{r}}1}
(\mathbf{B}_{1}), \bar{R}_{2{\rm{r}}} (\mathbf{D}_2)\}  \qquad \nonumber\\
\qquad +\min\{\hat{R}_{{\rm{r}}2} (\mathbf{B}_{2}\!), \bar{R}_{1{\rm{r}}}
(\mathbf{D}_1\!)\},
\end{eqnarray}
in which
\begin{eqnarray}
\bar{R}_{j{\rm{r}}}(\mathbf{D}_j) = \log{| \mathbf{I} +
(\mathbf{H}_{j{\rm{r}}} \mathbf{D}_j \mathbf{H}_{j{\rm{r}}}^{\rm
H}) (\sigma_{{\rm{r}}}^2)^{-1}|}, \label{e:Rbar}
\end{eqnarray}
and
\begin{eqnarray}
\hat{R}_{{\rm{r}}i}(\mathbf{B}_{i}) = \log|\mathbf{I} +
(\mathbf{H}_{{\rm{r}}i} \mathbf{B}_{i} \mathbf{H}_{{\rm{r}}i}^{\rm
H})(\sigma_{i}^2)^{-1}|.
\end{eqnarray}
For brevity of presentation, we define the following sum-rate of the BC phase
\begin{eqnarray}\label{e:RbcB}
R^{\rm bc}(\mathbf{B})=\hat{R}_{{\rm{r}}1} (\mathbf{B}_{1}) +
\hat{R}_{{\rm{r}}2} (\mathbf{B}_{2})
\end{eqnarray}
to represent the summation in the above equation hereafter.
}

For the relay optimization scenario considered here, the relay maximizes the sum-rate in \eqref{e:eqRtw} using minimum transmission power given the power allocation strategies of the
source nodes.\footnote{The term `sum-rate' by default means $R^{\rm tw}(\mathbf{B}, \mathbf{D})$ when we do not specify it to be the sum-rate of the BC or MA phase.} Since the relay needs to know $\mathbf{W}_1$ and $\mathbf{W}_2$ for decoding $\mathbf{s}_1$ and $\mathbf{s}_2$, respectively, as well as for designing $\mathbf{T}_{\mathrm{r}1}$ and $\mathbf{T}_{\mathrm{r}2}$, the source nodes should send their respective precoding matrices to the relay after they decide their transmit strategies. Similarly, the relay should also send $\mathbf{T}_{\mathrm{r}1}$ and $\mathbf{T}_{\mathrm{r}2}$ to both source nodes.

Given the above system model, we next solve the relay optimization problem.

\section{Relay optimization}\label{s:relayopt}
In the relay optimization scenario, the relay and the source nodes do not coordinate in choosing their respective power allocation strategies. Instead, the relay aims at maximizing $R^{\rm tw}(\mathbf{B}, \mathbf{D})$ in \eqref{e:eqRtw} with minimum power consumption after the source nodes decide their strategies and inform the relay.

Denote the power allocation that the source nodes decide to use as
$\mathbf{D}^0=[\mathbf{D}_1^0, \mathbf{D}_2^0]$.\footnote{The source nodes may determine their power allocation strategies using different objectives. Note that different source node power allocation strategies lead to different solutions of the relay optimization problem. However, the approach adopted for solving the relay optimization problem is valid for arbitrary source node power allocation.} For maximizing
the sum-rate given $\mathbf{D}^0$, the relay solves the following optimization problem\footnote{The positive semi-definite
constraints $\mathbf{D}_i\succeq0, \forall i$ and
$\mathbf{B}_i\succeq0, \forall i$ are assumed as default and
omitted for brevity in all formations of optimization problems in
this paper.}
\begin{subequations}\label{e:Rl}
\begin{align}
&\mathop{\mathbf{max}}\limits_{\mathbf{B}} \quad R^{\rm
tw}(\mathbf{B}, \mathbf{D}^0) \label{e:Rlobj}\\
&\;\;\mathbf{ s.t.} \quad\;\text{Tr}\{\mathbf{B}_{1}+\mathbf{B}_{2} \}\leq P_{\rm
r}^{\rm max}.\label{e:Rlcons}
\end{align}
\end{subequations}
The problem \eqref{e:Rl} is convex. However, in order to find the
optimal $\mathbf{B}$ with minimum
$\text{Tr}\{\mathbf{B}_{1}+\mathbf{B}_{2}\}$ among all possible
$\mathbf{B}$'s that achieve the same maximum of the objective
function in \eqref{e:Rl}, extra constraints need to be considered.
Two necessary constraints\footnote{These two necessary constraints are introduced here to show that the considered relay optimization problem is nonconvex. For the sufficient and necessary condition that a power allocation strategy is optimal in terms of maximizing sum-rate with minimum power consumption, please see Theorem~2.} are
\begin{subequations}\label{e:RlconsPM}
\begin{align}
\hat{R}_{{\rm r}i}(\mathbf{B}_{i})\leq \bar{R}_{j{\rm r}}
(\mathbf{D}_j^0), \forall i \label{e:RlconsPM1}\\
R(\mathbf{B}, \mathbf{D}^0)
\leq R^{\rm ma}(\mathbf{D}^0).\label{e:RlconsPM2}
\end{align}
\end{subequations}

\textcolor{Black}{
The constraint \eqref{e:RlconsPM1} is necessary because, due to the expression of $R(\mathbf{B}, \mathbf{D})$ in \eqref{e:RBD}, the power consumption of the relay can be reduced without decreasing the sum-rate $R^{\rm tw}(\mathbf{B}, \mathbf{D})$ in \eqref{e:eqRtw} given $\mathbf{D}^0$ by reducing $\text{Tr}\{\mathbf{B}_i\}$  if $\hat{R}_{\mathrm{r}i}(\mathbf{B}_{i})> \bar{R}_{j\mathrm{r}}(\mathbf{D}_j^0)$. Note that \eqref{e:RlconsPM1} is not necessarily satisfied with equality at optimality. In fact, it can be shown that \eqref{e:RlconsPM1} should be satisfied with inequality for at least one $i$ at optimality using  subsequent results in Section~\ref{s:MainBodyP2}. It can also be shown that \eqref{e:RlconsPM1} can be satisfied with inequalities for both $i$'s at optimality even if the relay has an unlimited power budget. We stress that \eqref{e:RlconsPM1} is not sufficient for obtaining the optimal solution. Other constraints are also needed including \eqref{e:RlconsPM2}. The constraint \eqref{e:RlconsPM2} is also necessary because if it is not satisfied given $\mathbf{D}^0$, then the power consumption of the relay can be reduced without decreasing the sum-rate $R^{\rm tw}(\mathbf{B}, \mathbf{D}^0)$ by decreasing $R(\mathbf{B}, \mathbf{D}^0)$ so that $R(\mathbf{B}, \mathbf{D}^0)
= R^{\rm ma}(\mathbf{D}^0)$.
}

The constraints in \eqref{e:RlconsPM} make the considered problem nonconvex. The
objective in this section is to find an efficient method of
deriving the optimal power allocation of the relay in the
considered scenario of relay optimization. It is straightforward
to see that the power allocation of the relay should be based on
waterfilling for relaying the signal in either direction regardless of
how the relay distributes its power between relaying the signals
in the two directions. This is due to the fact that the BC phase
is interference free since both source nodes are able to subtract
their self-interference. If the objective were to maximize $R^{\rm
bc}(\mathbf{B})$ instead of $R^{\rm tw}(\mathbf{B},
\mathbf{D}^0)$, the optimal strategy of the relay  could be found
via a simple search. Indeed, in that case, we could find the
optimal power allocation of the relay and consequently the optimal
$\mathbf{B}$ by searching for the optimal proportion that the
relay distributes its power between relaying the signals in the two
directions. However, such approach is infeasible for the
considered problem. The reason is that first of all it is unknown
what is the total power that the relay uses in the optimal
solution. As power efficiency is also considered, the relay may
not use full power in its optimal strategy. Moreover, from the
expression of $R^{\rm tw}(\mathbf{B}, \mathbf{D})$ in
\eqref{e:eqRtw}, it can be seen that the maximum achievable
$R^{\rm tw}(\mathbf{B}, \mathbf{D}^0)$ also depends on
$\bar{R}_{1{\rm r}}(\mathbf{D}_1^0)$, $\bar{R}_{2{\rm
r}}(\mathbf{D}_2^0)$, and $R^{\rm ma}(\mathbf{D}^0)$. Due to this
dependence, the two constraints in \eqref{e:RlconsPM} are
necessary for the considered problem of sum-rate maximization with minimum power consumption. However, these two
constraints are implicit in the sense that they are constraints on
the rates instead of on the power allocation of the relay. Such
constraints offer no insight in finding the optimal $\mathbf{B}$.
In order to transform the above mentioned dependence of $R^{\rm
tw}(\mathbf{B}, \mathbf{D}^0)$ on $\bar{R}_{1{\rm
r}}(\mathbf{D}_1^0)$, $\bar{R}_{2{\rm r}}(\mathbf{D}_2^0)$, and
$R^{\rm ma}(\mathbf{D}^0)$ into an explicit form, and to discover the
insight behind the constraints in \eqref{e:RlconsPM}, we next propose
the idea of relative water-levels and develop a method based on this idea.

\subsection{Relative water-levels}
Denote the rank of $\mathbf{H}_{{\rm r}i}$ as $\mathrm{r}_{{\rm
r}i}$ and the singular value decomposition (SVD) of
$\mathbf{H}_{{\rm r}i}$ as $\mathbf{U}_{{\rm r}i}
\mathbf{\Omega}_{{\rm r}i}\mathbf{V}_{{\rm r}i}^\mathrm{H}$.
Assume that the first $\mathrm{r}_{{\rm r}i}$ diagonal elements of
$\mathbf{\Omega}_{{\rm r}i}$ are non-zero, sorted in descending
order and denoted as $\omega_{{\rm r}i}(1),\dots, \omega_{\mathrm{r}i}({\rm
r}_{{\rm r}i})$, while the last $\min\{n_i, n_{\rm r}\}-r_{{\rm
r}i}$ diagonal elements are zeros. Define $\mathcal{I}_i=\{1,
\dots, {\rm r}_{{\rm r}i}\}, \forall i$ and $\alpha_i(k) =
{|\omega_{{\rm r}i}(k)|^2}/{\sigma_i^2}, \forall
k\in\mathcal{I}_i, \forall i$. For a given $\mathbf{D}=[\mathbf{D}_1, \mathbf{D}_2]$, define
$\mu_1(\mathbf{D}_1)$, $\mu_2(\mathbf{D}_2)$, and $\mu_{\rm
ma}(\mathbf{D})$ such that
\begin{subequations}
\begin{align}
\hspace{-2mm}&\sum\limits_{k\in\mathcal{I}_2} \log\bigg(1+\big(
\frac{1}{\mu_1(\mathbf{D}_1)}\alpha_2(k)-1\big)^{+}\bigg) =
\bar{R}_{1{\rm r}}(\mathbf{D}_1)\label{e:watermu1}\\
\hspace{-2mm}&\sum\limits_{k\in\mathcal{I}_1} \log\bigg(1 +
\big(\frac{1}{\mu_2(\mathbf{D}_2)}\alpha_1(k)-1\big)^{+}
\bigg)=\bar{R}_{2{\rm r}}(\mathbf{D}_2)\label{e:watermu2} \\
\hspace{-2mm}&\sum\limits_{i}\!\sum\limits_{k\in\mathcal{I}_i}
\!\log\bigg(\!1+\big(\frac{1}{\mu_{\rm ma}(\mathbf{D})}
\alpha_i(k)-1\big)^{+}\!\bigg)\!=R^{\rm
ma}(\mathbf{D})\label{e:watermum}
\end{align}
\end{subequations}
where $(\cdot)^{+}$ stands for projection to the positive orthant.
The physical meaning of $\mu_i(\mathbf{D}_i)$ is that
if waterfilling is performed on $\omega_{\mathrm{r}j}(k)$'s, $\forall
k\in \mathcal{I}_j$ using the water-level $1/\mu_i(\mathbf{D}_i)$,
then the information rate of the transmission from the relay to
source node $j$ using the resulting waterfilling-based power
allocation achieves precisely $\bar{R}_{i\mathrm{r}}
(\mathbf{D}_i)$. The physical meaning of $\mu_{\rm
ma}(\mathbf{D})$ is that if waterfilling is performed on
$\omega_{\mathrm{r}i}(k)$'s, $\forall k\in \mathcal{I}_i, \forall i$
using the water-level $1/\mu_{\rm ma}(\mathbf{D})$, then the
sum-rate of the transmission from the relay to the two source
nodes using the resulting waterfilling-based power allocation
achieves precisely ${R}^{\rm ma}(\mathbf{D})$. Note that
$1/\mu_i(\mathbf{D}_i), \forall i$ and $1/\mu_{\rm m}(\mathbf{D})$
are not the actual water-levels for the MA or the BC phase. They
are just relative water-levels introduced to transfer and simplify
the constraints in \eqref{e:RlconsPM}. Denote the actual
water-levels used by the relay for relaying the signal from source
node $j$ to source node $i$ as $1/\lambda_i, \forall i$. With
water-level $1/\lambda_i$, $\mathbf{B}_i$ can be given as
$\mathbf{B}_i=\mathbf{V}_{\mathrm{r}i}\mathbf{P}_{\mathrm{r}i}
(\lambda_i)\mathbf{V}_{\mathrm{r}i}^\mathrm{H}$ where 
$\mathbf{P}_{\mathrm{r}i}(\lambda_i)=\text{diag}\bigg(\big(\frac{1}{\lambda_i} - \frac{1}{\alpha_i(1)}\big)^{+}, \dots, \big(\frac{1}{\lambda_i}\!-\!\frac{1}{\alpha_i(\mathrm{r}_{\mathrm{r}i})}\big)^{+}, 0,\dots, 0\bigg)$
in which $\text{diag}(\cdot)$ stands for making a diagonal matrix using the given elements, $(\cdot)^{+}$ stands for projection to the positive orthant, and $\mathbf{0}_{n_\mathrm{r}-\mathrm{r}_{\mathrm{r}i}}$
stands for all-zero matrix of size $(n_\mathrm{r} -
\mathrm{r}_{\mathrm{r}i}) \times (n_\mathrm{r} -
\mathrm{r}_{\mathrm{r}i})$. \textcolor{Black}{The power
allocated on $\omega_{\mathrm{r}i}(k)$ is $p_{\mathrm{r}i}(k)=\big({1} /
{\lambda_i}-{1}/{\alpha_i(k)}\big)^{+}, \forall k\in \mathcal{I}_i,
\forall i$}. The resulting rate
$\hat{R}_{\mathrm{r}i} (\mathbf{B}_{i})$ is given by
$\sum\limits_{k\in\mathcal{I}_{i}} \!\log\!\bigg( \!1\!+\!\big(
{\alpha_i(k)}/{\lambda_i} -\!1\big)^{+}\!\bigg)$. Using
$\mu_1(\mathbf{D}_1)$, $\mu_2(\mathbf{D}_2)$, and $\mu_{\rm
ma}(\mathbf{D})$, the constraints in \eqref{e:RlconsPM1} can be
rewritten as
\begin{subequations}\label{e:WLcons}
\begin{align}
\lambda_i \geq \mu_j(\mathbf{D}_j^0), \forall i \quad\quad\quad
\quad\quad\quad\;\,
\label{e:WLcons1} \\
\sum\limits_{i}\sum\limits_{k\in\mathcal{I}_{i}} \!\log\! \bigg(
\!1\!+\!\big(\frac{1}{\lambda_i}\alpha_i(k)-\!1 \big)^{+}
\!\bigg) \quad\quad\quad\quad\quad\quad\quad\;\, \nonumber\\
\leq \sum\limits_{i} \sum\limits_{k\in\mathcal{I}_{i}}
\!\log\!\bigg(\!1\! +\!\big( \frac{1}{\mu_\mathrm{ma}
(\mathbf{D}^0)}\alpha_i(k) - \!1\big)^{+} \!\bigg).\label{e:WLcons2}
\end{align}
\end{subequations}

\textcolor{Black}{
Given \eqref{e:watermu1} and \eqref{e:watermu2}, it is not difficult to see that \eqref{e:RlconsPM1} is equivalent to \eqref{e:WLcons1}. Moreover, the equivalence between \eqref{e:RlconsPM2} and \eqref{e:WLcons2} can be explained as follows. Given $\mathbf{D}^0$ and \eqref{e:RlconsPM2}, $R^{\rm tw}(\mathbf{B}, \mathbf{D}^0)$ in \eqref{e:Rlobj} becomes $R (\mathbf{B},\mathbf{D}^0)/2$. Given \eqref{e:RlconsPM1}, or equivalently \eqref{e:WLcons1}, $R (\mathbf{B},\mathbf{D})$ in \eqref{e:RBD} with $\mathbf{D}=\mathbf{D}^0$ becomes $\hat{R}_{{\rm{r}}1} (\mathbf{B}_{1})+\hat{R}_{{\rm{r}}2} (\mathbf{B}_{2})$. Then, substituting the left-hand side of \eqref{e:RlconsPM2} with $\hat{R}_{{\rm{r}}1} (\mathbf{B}_{1})+\hat{R}_{{\rm{r}}2} (\mathbf{B}_{2})$, i.e., $R^{\rm bc}(\mathbf{B})$ in \eqref{e:RbcB}, and using \eqref{e:watermum}, the constraint \eqref{e:WLcons2} is obtained.
}

The procedure for the relay optimization can be summarized in the
following three steps:

1. Obtain $\mu_1(\mathbf{D}_1^0)$, $\mu_2(\mathbf{D}_2^0)$, and
$\mu_\mathrm{ma}(\mathbf{D}^0)$ from $\mathbf{D}^0$;

2. Determine the optimal $\lambda_i$;

3. Obtain $\mathbf{P}_{\mathrm{r}i}(\lambda_i)$ and $\mathbf{B}_i$
from $\lambda_i$.

The first and the third steps are straightforward given the
definitions \eqref{e:watermu1}-\eqref{e:watermum} and
\eqref{e:WL2PW}. Therefore, finding the optimal $\lambda_i,
\forall i$ in the second step is the essential part to be dealt
with later in this section.

From hereon, $\mu_1(\mathbf{D}_1)$, $\mu_2(\mathbf{D}_2)$, and
$\mu_{\rm ma}(\mathbf{D})$ are denoted as $\mu_1$, $\mu_2$ and
$\mu_\mathrm{ma}$, respectively, for brevity. The same
markers/superscripts on $\mathbf{D}_i$ and/or $\mathbf{D}$ are
used on $\mu_i$ and/or $\mu_\mathrm{ma}$ to represent the
connection. For example, $\mu_i(\mathbf{D}_i^{0})$ and
$\mu_\mathrm{ma}(\tilde{\mathbf{D}})$ are briefly denoted as
$\mu_i^{0}$ and $\tilde{\mu}_\mathrm{ma}$, respectively. The rate
$\hat{R}_{\mathrm{r}i}(\mathbf{B}_{i})$ obtained using water-level
$1/\lambda_i$ is also denoted as
$\hat{R}_{\mathrm{r}i}(\lambda_{i})$.

\subsection{Algorithm for relay optimization}\label{s:MainBodyP2}

Using the relative water-levels $\mu_i,\forall i$ and $\mu_{\rm ma}$, we can now develop the algorithm for relay optimization. In order to do that, the following lemmas are presented.

\emph{Lemma 1}: $1/\mu_{\rm ma} <\max\{1/\mu_1, 1/\mu_2\}$.

\textbf{Proof}: The proof for Lemma~1 is straightforward. Using
\eqref{e:watermu1}-\eqref{e:watermum}, it can be seen that $R^{\rm
ma}(\mathbf{D})\geq \sum\limits_i \bar{R}_{i\mathrm{r}}
(\mathbf{D}_i)$ if $1/\mu_{\rm ma}\geq \max\{1/\mu_1, 1/\mu_2\}$.
However, \textcolor{Black}{given the definitions in \eqref{e:eqmacR}
and \eqref{e:Rbar}, it can be seen that $R^{\rm ma}(\mathbf{D})\geq
\sum\limits_i \bar{R}_{i\mathrm{r}}(\mathbf{D}_i)$ is impossible
\cite{Mimo_Cap}. Therefore, $1/\mu_{\rm ma}<\max\{1/\mu_1, 1/\mu_2\}$. \hfill$\blacksquare$
}

\emph{Lemma 2}: Assume that there exist $\{\lambda_i, \lambda_j\}$
and $\{\lambda_i^{\prime}, \lambda_j^{\prime}\}$ such that
$\lambda_i^{\prime}<\lambda_i\leq\lambda_j<\lambda_j^{\prime}$. If
$\sum\limits_{l} \text{Tr} \{ \mathbf{P}_{\mathrm{r}l}
(\lambda_l)\}=\sum\limits_{l} \text{Tr}\{ \mathbf{P}_{\mathrm{r}l}
(\lambda_l^{\prime})\}$, then $\sum\limits_l \hat{R}_{\mathrm{r}l}
(\lambda_{l})>\sum\limits_l \hat{R}_{\mathrm{r}l}
(\lambda_{l}^{\prime})$ as long as $1/\lambda_j
> \min\limits_k\{1 / \alpha_j(k)\}$.

\textbf{Proof}: See Subsection \ref{s:PLm2} in Appendix.
\hfill$\blacksquare$

\textcolor{Black}{
Lemma~2 states that, for any given $\{\lambda_1, \lambda_2\}$ such that $1/\lambda_2
> \min\limits_k\{1 / \alpha_2(k)\}$ assuming $\lambda_1\leq\lambda_2$, decreasing $\min\{\lambda_1, \lambda_2\}$
and increasing $\max\{\lambda_1, \lambda_2\}$ while fixing the total power
consumption leads to a smaller BC phase sum-rate than that achieved by using
$\{\lambda_1, \lambda_2\}$.
}

\emph{Lemma 3}: Assume that there exist $\{\lambda_i, \lambda_j\}$
and $\{\lambda_i^{\prime}, \lambda_j^{\prime}\}$ such that
$\lambda_i<\lambda_j$, $\lambda_i^{\prime}>\lambda_i$ and
$\lambda_j^{\prime}>\lambda_j$, and
\begin{eqnarray}
\hat{R}_{\mathrm{r}i}(\lambda_{i}^{\prime})+ \hat{R}_{\mathrm{r}j}
(\lambda_{j}) = \hat{R}_{\mathrm{r}i}(\lambda_{i})+
\hat{R}_{\mathrm{r}j} (\lambda_{j}^{\prime})
\end{eqnarray}
then as long as $\lambda_i^{\prime} \leq \lambda_j$, it holds true
that
\begin{equation}
\text{Tr} \{ \mathbf{P}_{\mathrm{r}i}(\lambda_i^{\prime})\} +
\text{Tr} \{ \mathbf{P}_{\mathrm{r}j}(\lambda_j)\} < \text{Tr} \{
\mathbf{P}_{\mathrm{r}i} (\lambda_i)\} + \text{Tr} \{
\mathbf{P}_{\mathrm{r}j} (\lambda_j^{\prime})\}.
\end{equation}

\textbf{Proof}: See Subsection \ref{s:PLm3} in Appendix.
\hfill$\blacksquare$

\textcolor{Black}{
Lemma~3 states that, for any given $\{\lambda_1, \lambda_2\}$, decreasing $\min\{\lambda_1, \lambda_2\}$
and increasing $\max\{\lambda_1, \lambda_2\}$ such that the BC phase sum-rate is unchanged, the power
consumption increases.
}

\textbf{Theorem 1}: The optimal solution of the considered relay optimization problem always satisfies the following properties
\begin{subequations}
\begin{align}
&\min\bigg\{\frac{1}{\lambda_1}, \frac{1}{\lambda_2}\bigg\}= \min \bigg\{\frac{1}{\mu_1^0}, \frac{1}{\mu_2^0}\bigg\} \quad \text{if} \quad \lambda_1\neq \lambda_2 \; \label{e:optpropty1}\\
&\frac{1}{\lambda_1}=\frac{1}{\lambda_2}= \min \bigg\{\frac{1}{\mu_{\rm ma}^0}, \frac{1}{\lambda^0}\bigg\} \quad \text{if} \quad \lambda_1= \lambda_2 \qquad\label{e:optpropty2}
\end{align}
\end{subequations}
in which ${1}/{\lambda^0}$ is the water-level obtained by waterfilling $P_\mathrm{r}^{\rm max}$ on $\omega_{\mathrm{r}i}(k), \forall k\in \mathcal{I}_i, \forall i$.

\textbf{Proof}: See Subsection \ref{s:PTh1} in Appendix.
\hfill$\blacksquare$

\textcolor{Black}{According to the proof of Theorem~1, it can be seen that $\lambda_1\neq \lambda_2$ at optimality and consequently the equation in \eqref{e:optpropty1} holds when both of the following two conditions are satisfied: (i) the relay has sufficient power, i.e., $1/\lambda^0>\min\{1/\mu_1^0,  1/\mu_2^0\}$, and (ii) there is asymmetry between $\mu_1^0$ and $\mu_2^0$, i.e., $\min\{1/\mu_1^0, 1/\mu_2^0\}<1/\mu_{\rm ma}^0<\max\{1/\mu_1^0, 1/\mu_2^0\}$. If either of the above two conditions is not satisfied, $\lambda_1= \lambda_2$ at optimality and consequently the equation in \eqref{e:optpropty2} holds. }

\textbf{Theorem 2}: The conditions \eqref{e:WLcons1}, \eqref{e:WLcons2}, \eqref{e:optpropty1}, and \eqref{e:optpropty2} are sufficient and necessary to determine the optimal $\{\lambda_1, \lambda_2\}$ with minimum power consumption for the relay optimization problem among all $\{\lambda_1, \lambda_2\}$'s that maximize the sum-rate $R^{\rm tw}(\mathbf{B}, \mathbf{D}^0)$.

\textbf{Proof}: See Subsection \ref{s:PTh2} in Appendix.
\hfill$\blacksquare$

\textcolor{Black}
{
It should be noted that the power constraint \eqref{e:Rlcons} is not always tight at optimality due to the constraints in \eqref{e:WLcons1}, \eqref{e:WLcons2} (or equivalently \eqref{e:RlconsPM1}, \eqref{e:RlconsPM2}), \eqref{e:optpropty1}, and \eqref{e:optpropty2}. Each of \eqref{e:WLcons1}, \eqref{e:WLcons2}, \eqref{e:optpropty1}, and \eqref{e:optpropty2} may refrain the relay from using its full power at optimality. The reason can be found from the proofs of Theorems~1~and~2. Specifically, \eqref{e:WLcons1} and \eqref{e:optpropty1} make sure that there is no superfluous power spent for relaying the signal in each direction while \eqref{e:WLcons2} and \eqref{e:optpropty2} guarantee that the power consumption of the relay cannot be further reduced without reducing the sum-rate.
}

\begin{table}
\begin{center}
\caption {The algorithm for relay optimization.}\label{t:RPA}
\begin{tabular*}{0.48\textwidth}[]{p{0.46\textwidth}}
\hline\hline 
1. Initial waterfilling: allocate $P_\mathrm{r}^{\rm max}$ on
$\omega_{\mathrm{r}i}(k), \forall k\in \mathcal{I}_i, \forall i$
using waterfilling. Denote the initial water level as ${1}/{\lambda^{0}}$.
Set ${1}/{\lambda_{1}}={1}/{\lambda_{2}}={1}/{\lambda^{0}}$. The power
allocated on $\omega_{\mathrm{r}i}(k)$ is $p_{\mathrm{r}i}(k)=\big({1} /
{\lambda_i}-{1}/{\alpha_i(k)}\big)^{+}, \forall k\in \mathcal{I}_i,
\forall i$. \\
\hline
2. Check if ${1}/{\lambda_{i}}\leq 1/\mu_j^0$ for both  $i=1, 2$. If yes,
proceed to Step~6. Otherwise, assume that ${1}/{\lambda_1}> 1/\mu_2^0$, proceed to Step~3.\\
\hline
3. Set $\lambda_1=\mu_2^0$. Check if $1/\lambda_2 <
1/\mu_1^0$. If not, proceed to Step~4. Otherwise, proceed to
Step~5. \\
\hline
4. Calculate $P_\mathrm{r}^{\prime} = P_\mathrm{r}^{\rm max} -
\sum\limits_{k\in\mathcal{I}_1}{p_{\mathrm{r}1}(k)}$. Allocate $P_\mathrm{r}^{\prime}$
on $\omega_{\mathrm{r}2}(k)$'s,$\forall k\in \mathcal{I}_2$ via waterfilling.
Obtain the water level $1/\lambda_2$. If $1/\lambda_2> 1/\mu_1^0$,
proceed to Step~5. Otherwise, go to Step~6. \\
\hline
5. Set $\lambda_2=\mu_1^0$ and proceed to Step~6.\\
\hline
6. If $1/\lambda_i\geq 1/\mu_\mathrm{ma}^0, \forall i$, set
$\lambda_i = \mu_\mathrm{ma}^0, \forall i$. Check if
$1/\lambda_i\leq 1/\mu_\mathrm{ma}^0, \forall i$. If yes, output
$\lambda_i, \forall i$ and break. Otherwise, check if
$\sum\limits_i\hat{R}_{\mathrm{r}i}(\lambda_{i}) \leq R^{\rm ma}
(\mathbf{D}^0)$. If yes, output $\lambda_i, \forall i$ and break.
Otherwise, proceed to Step~7. \\
\hline 7. Assuming that $\lambda_j<\lambda_i$, find
$\lambda_{j}^{\prime}$ such that $|\mathcal{M}_{\mathrm{r}j}^{+} |
\log\lambda_j^{\prime} = \sum\limits_{k\in\mathcal{M}_{\mathrm{r}j}^{+}}
\log\alpha_j(k)-R^{\rm ma}(\mathbf{D}^0)+\bar{R}_{j\mathrm{r}}
(\mathbf{D}_j^0)$, where $p_{\mathrm{r}j}(k)=\big({1}/{\lambda_j^{\prime}}
-{1}/{\alpha_j(k)}\big)^{+}, \forall k\in \mathcal{I}_j$,
$\mathcal{M}_{\mathrm{r}j}^{+}=\{k|p_{\mathrm{r}j}(k)>0\}$ and
$|\mathcal{M}_{\mathrm{r}j}^{+}|$ is the cardinality of the set
$\mathcal{M}_{\mathrm{r}j}^{+}$. Set $\lambda_j=\lambda_j^{\prime}$
and output $\lambda_i$ and $\lambda_j$. \\
\hline \hline
\end{tabular*}
\end{center}
\vspace{-0.5cm}
\end{table}

Based on the above results in Theorem~1 and Theorem~2, the algorithm summarized in Table~\ref{t:RPA} is proposed to find the optimal relay power allocation for the relay optimization problem. The algorithm can be briefly understood as
follows. Step~1 performs initial power allocation and obtains the
initial water level $\lambda^0$. The water-levels $\lambda_i =
\lambda^0, \forall i$ maximize $R^{\rm bc}(\mathbf{B})$ among all
possible $\{\lambda_1, \lambda_2\}$ combinations subject to the
power limit of the relay. Step~2 checks if $\min\{\hat{R}_{{\rm{r}}i}
(\mathbf{B}_{i}), \bar{R}_{j{\rm{r}}} (\mathbf{D}_j)\}$ is
upper-bounded by $\bar{R}_{j\mathrm{r}}(\mathbf{D}_j^0), \forall
i$. If $\hat{R}_{\mathrm{r}1}(\lambda_{1}^{0}) > \bar{R}_{2
\mathrm{r}} (\mathbf{D}_2^0)$, the relay reduces its transmission
power allocated for relaying the signal from source node $2$ to
source node $1$ so that $\hat{R}_{\mathrm{r}1}(\lambda_{1})=
\bar{R}_{2\mathrm{r}} (\mathbf{D}_2^0)$ in Step~3. In the case
that $\hat{R}_{\mathrm{r}1}(\lambda_{1})$ is reduced in Step~3, in terms of increasing $\lambda_{1}$,
extra power becomes available for relaying the signal from source
node $1$ to source node $2$. Therefore, if $\hat{R}_{\mathrm{r}2}
(\lambda_{2}^{0})<\bar{R}_{1\mathrm{r}} (\mathbf{D}_1^0)$, the
remaining power of the relay is allocated for relaying the signal
from source node $1$ to source node $2$ at first in
Step~4. Later in Step~4, it is checked if
$\hat{R}_{\mathrm{r}2}(\lambda_{2})>\bar{R}_{1\mathrm{r}}
(\mathbf{D}_1^0)$ under the new power allocation. If
$\hat{R}_{\mathrm{r}2} (\lambda_{2}) > \bar{R}_{1 \mathrm{r}}
(\mathbf{D}_1^0)$ in Step~4, the relay reduces its
transmission power allocated for relaying the signal from source
node $1$ to source node $2$ so that $\hat{R}_{\mathrm{r}2}
(\lambda_{2})= \bar{R}_{1\mathrm{r}}(\mathbf{D}_1^0)$ in Step~5.
Steps~6 checks if $\hat{R}_{\mathrm{r}1} (\lambda_{1}) +
\hat{R}_{\mathrm{r}2}(\lambda_2)\leq R^{\rm ma}(\mathbf{D}^0)$.
In the case that this constraint is not satisfied, Step~6 or
Step~7 revise the power allocation so that $\hat{R}_{\mathrm{r}1}
(\lambda_{1}) + \hat{R}_{\mathrm{r}2}(\lambda_2)= R^{\rm
ma}(\mathbf{D}^0)$ and the power consumption of the relay is
minimized. The above procedure in the proposed algorithm, which
terminates after Step~6~or~7, is not iterative.

The following theorem regarding the proposed algorithm is in order.

\textbf{Theorem 3}: The water-levels obtained using the algorithm
for relay optimization in Table~\ref{t:RPA} achieve the optimal relay
power allocation for the considered relay optimization problem of sum-rate maximization with minimum relay power consumption.

\textbf{Proof}: See Subsection \ref{s:PTh3} in Appendix.
\hfill$\blacksquare$

Depending on the source node power allocation strategies and the
power limit at the relay, different results can be obtained at the
output of the algorithm in Table~\ref{t:RPA}. Define the power
thresholds $P_\mathrm{ma}=\sum\limits_i\sum\limits_{k\in\mathcal{I}_i}
\big({1}/{\mu_\mathrm{ma}^0}-{1}/{\alpha_i(k)}\big)^{+}$,
$P_\mathrm{l} = \sum\limits_i\sum\limits_{k\in\mathcal{I}_i} \big({1}/{\max\{\mu_1^0,
\mu_2^0\}}-{1}/{\alpha_i(k)}\big)^{+}$,
$P_\mathrm{t}=\sum\limits_i\sum\limits_{k\in\mathcal{I}_i} \big({1}/
{\mu_i^0}-{1} / {\alpha_i(k)}\big)^{+}$ and $P_\mathrm{s} =
\sum\limits_i \sum\limits_{k\in\mathcal{I}_i} \big({1}/\min\{\mu_1^0,
\mu_2^0\}-{1}/{\alpha_i(k)}\big)^{+}$. Recall from Lemma~1 that
$\mu_{\rm ma}^0> \min\{\mu_1^0, \mu_2^0\}$.

For the case that
$\mu_{\rm ma}^0 \geq \max\{\mu_1^0, \mu_2^0\}$, the following
subcases exit as $P_\mathrm{r}^\mathrm{max}$ increases. If
$P_\mathrm{r}^\mathrm{max}$ is small such that
$P_\mathrm{r}^\mathrm{\rm max} < P_\mathrm{\rm ma}$, the algorithm
proceeds through Steps~1-2-6 and
\begin{subequations}
\begin{align}
&\quad\;\lambda_i=\lambda^0> \mu_\mathrm{ma}^0, \forall i \label{e:C1s1e1}\\
&\sum\limits_{i} \text{Tr}\{\mathbf{P}_{\mathrm{r}i}(\lambda_i)\} =
P_\mathrm{r}^\mathrm{max}
\end{align}
\end{subequations}
at the output of the algorithm, while \eqref{e:WLcons1} and
\eqref{e:WLcons2} are satisfied with inequality. Note that some power of the source nodes is wasted in this subcase. Since the sum-rate $R^{\rm tw}(\mathbf{B}, \mathbf{D})$ is bounded by $\hat{R}_{\mathrm{r}1} (\lambda_{1})+\hat{R}_{\mathrm{r}2}(\lambda_2)$ due to the small power limit of the relay, the source nodes could use less power without reducing $R^{\rm tw}(\mathbf{B}, \mathbf{D})$ if there would be coordination in the system. Indeed, if the source nodes could be coordinated to optimize their power allocation as well, they only need to use the power of $\text{Tr}\{\mathbf{D}^\dag_1\}$+$\text{Tr}\{\mathbf{D}^\dag_2\}$, where $\mathbf{D}^\dag=[\mathbf{D}^\dag_1, \mathbf{D}^\dag_2]$ is the optimal solution to the following problem
\begin{subequations}\label{e:SubIfnl}
\begin{align}
&\mathop{\mathbf{min}}\limits_{\mathbf{D}}\; \text{Tr}\{\mathbf{D}_1\}+\text{Tr}\{\mathbf{D}_2\} \\
&\mathbf{\;\;s.t.}  \; R^{\rm ma}(\mathbf{D})\geq \hat{R}_{\mathrm{r}1} (\lambda^0)+\hat{R}_{\mathrm{r}2}(\lambda^0) \\
&\qquad\;\bar{R}_{1\mathrm{r}}(\mathbf{D}_1)\geq \hat{R}_{\mathrm{r}2}(\lambda^0)\\
&\qquad\;\bar{R}_{2\mathrm{r}}(\mathbf{D}_2)\geq \hat{R}_{\mathrm{r}1}(\lambda^0). 
\end{align}
\end{subequations}
It can be shown that $\text{Tr}\{\mathbf{D}^0_1\}+\text{Tr}\{\mathbf{D}^0_2\} >\text{Tr}\{\mathbf{D}^\dag_1\}+\text{Tr}\{\mathbf{D}^\dag_2\}$ in this subcase. Therefore, the power of $\text{Tr}\{\mathbf{D}^0_1\}+\text{Tr}\{\mathbf{D}^0_2\}-\text{Tr}\{\mathbf{D}^\dag_1\}-\text{Tr}\{\mathbf{D}^\dag_2\}$ is wasted at the source nodes because of the lack of coordination. 

Increasing
$P_\mathrm{r}^\mathrm{max}$ such that $P_\mathrm{ma} \leq
P_\mathrm{r}^\mathrm{max}\leq P_\mathrm{l}$,  the algorithm
proceeds through Steps~1-2-6. Increasing
$P_\mathrm{r}^\mathrm{max}$ such that $P_\mathrm{l}
<P_\mathrm{r}^\mathrm{max}\leq P_\mathrm{t}$, the algorithm
proceeds through Steps~1-2-3-4-6. Further increasing
$P_\mathrm{r}^\mathrm{max}$ such that $P_\mathrm{t}
<P_\mathrm{r}^\mathrm{max}\leq P_\mathrm{s}$, the algorithm
proceeds through Steps~1-2-3-4-5-6. Further increasing
$P_\mathrm{r}^\mathrm{max}$ such that $P_\mathrm{r}^\mathrm{max} >
P_\mathrm{s}$, the algorithm proceeds through Steps~1-2-3-5-6. In the above subcases, it holds
that
\begin{subequations}
\begin{align}
&\quad\;\lambda_i= \mu_\mathrm{ma}^0 \geq \lambda^0 , \forall i \label{e:C1s2e1}\\
&\sum\limits_{i}\text{Tr}\{\mathbf{P}_{\mathrm{r}i}(\lambda_i)\} \leq
P_\mathrm{r}^\mathrm{max}
\end{align}
\end{subequations}
at the output of the algorithm, while \eqref{e:WLcons1} is
satisfied with inequality for each $i$ such that $1/\mu_i^0 >
1/\mu_\mathrm{ma}^0$ and \eqref{e:WLcons2} is satisfied with
equality. For these subcases, the sum-rate $R^{\rm tw}(\mathbf{B}, \mathbf{D})$ is bounded by  $R^{\rm ma}(\mathbf{D}^0)$ and there is no waste of power at the source nodes.

For the case that $\mu_\mathrm{ma}^0 < \max\{\mu_1^0, \mu_2^0\}$,
it holds that $\min\{\mu_1^0, \mu_2^0\}<\mu_\mathrm{ma}^0 <
\max\{\mu_1^0, \mu_2^0\}$ according to Lemma~1. Assume that
$\mu_2^0 > \mu_1^0$ and find $\bar{\lambda}_2$ such that
$\hat{R}_{\mathrm{r}2}(\bar{\lambda}_2)= R^{\rm ma}(\mathbf{D}^0)-
\bar{R}_{2 \mathrm{r}}(\mathbf{D}_2^0)$. Let $\bar{\lambda}_1
=\mu_2^0$ and define $\bar{P}_{\mathrm{ma}}=\sum\limits_i\sum\limits_{k\in\mathcal{I}_i} \big(1/\bar{\lambda}_i-{1} / {\alpha_i(k)}\big)^{+}$.
It can be seen from Lemma~3 that $\bar{P}_{\mathrm{ma}}>
P_{\mathrm{ma}}$. The following subcases appear as
$P_\mathrm{r}^\mathrm{max}$ increases. If
$P_\mathrm{r}^\mathrm{max}$ is small such that
$P_\mathrm{r}^\mathrm{max} < P_\mathrm{l}$, the algorithm proceeds
through Steps~1-2-6 and
\begin{subequations}
\begin{align}
&\lambda_i=\lambda^0> \max\{\mu_1^0, \mu_2^0\}, \forall i\\
&\;\sum\limits_{i} \text{Tr} \{ \mathbf{P}_{
\mathrm{r}i} (\lambda_i)\} =P_\mathrm{r}^\mathrm{max}
\end{align}
\end{subequations}
at the output of the algorithm, while \eqref{e:WLcons1} and
\eqref{e:WLcons2} are satisfied with inequality. Increasing
$P_\mathrm{r}^\mathrm{max}$ such that $P_\mathrm{l}\leq
P_\mathrm{r}^\mathrm{max} \leq \bar{P}_\mathrm{ma}$,  the
algorithm proceeds through Steps~1-2-3-4-6 and
\begin{subequations}
\begin{align}
&\quad\;\;\,\,\lambda_1=\mu_2^0\geq \lambda^0\\
&\sum\limits_{i}\text{Tr}\{\mathbf{P}_{\mathrm{r}i} (\lambda_i)\} =
P_\mathrm{r}^\mathrm{max}
\end{align}
\end{subequations}
at the output of the algorithm, while \eqref{e:WLcons1} is
satisfied with equality for $i=1$ and inequality for $i=2$. Note that there is waste of power at the source nodes for the above two subcases as long as $P_\mathrm{r}^\mathrm{max} <\bar{P}_\mathrm{ma}$ because the sum-rate $R^{\rm tw}(\mathbf{B}, \mathbf{D})$ is bounded by  $\hat{R}_{\mathrm{r}1} (\lambda_{1})+\hat{R}_{\mathrm{r}2}(\lambda_2)$.

Increasing $P_\mathrm{r}^\mathrm{max}$ such that
$\bar{P}_\mathrm{ma} <  P_\mathrm{r}^\mathrm{max} \leq
P_\mathrm{t}$, the algorithm proceeds through Steps~1-2-3-4-6-7.
Further increasing $P_\mathrm{r}^\mathrm{max}$ such that
$P_\mathrm{t}<P_\mathrm{r}^\mathrm{max} \leq P_\mathrm{s}$, the
algorithm proceeds through Steps~1-2-3-4-5-6-7. Further increasing
$P_\mathrm{r}^\mathrm{max}$ such that $P_\mathrm{r}^\mathrm{max} >
P_\mathrm{s}$, the algorithm proceeds through Steps~1-2-3-5-6-7.
In the subcases when $P_\mathrm{r}^\mathrm{max} \geq \bar{P}_\mathrm{ma}$,
it holds that
\begin{subequations}
\begin{align}
&\quad\;\;\,\,\lambda_1=\mu_2^0> \lambda^0\\
&\sum\limits_{i}
\text{Tr}\{\mathbf{P}_{\mathrm{r}i} (\lambda_i) \} \leq
P_\mathrm{r}^\mathrm{max}
\end{align}
\end{subequations}
at the output of the algorithm, while \eqref{e:WLcons1} is
satisfied with equality for $i=1$ and inequality for $i=2$, and
\eqref{e:WLcons2} is satisfied with equality. The optimal
$\lambda_2$ is found in Step~7 of the proposed algorithm.
For these subcases, there is no waste of power at the source nodes.

\begin{figure}[!t]
\psfrag{mu1}{\small$\frac{1}{\mu_1^0}$}
\psfrag{mu2}{\small$\frac{1}{\mu_2^0}$}
\psfrag{lbd0}{\small$\textcolor{red}{\frac{1}{\lambda^0}}$}
\psfrag{mum}{\small$\frac{1}{\mu_\mathrm{ma}^0}$}
\psfrag{mut}{\small$\frac{1}{\mu_\mathrm{t}^0}$}
\psfrag{blbd1}{\small(\!$\frac{1}{\bar{\lambda}_1}$\!)}
\psfrag{blbd2}{\small$\frac{1}{\bar{\lambda}_2}$}
\psfrag{aaa}{\tiny$\omega_{\mathrm{r}1}(1)$}
\psfrag{bbb}{\tiny$\omega_{\mathrm{r}1}(2)$}
\psfrag{ccc}{\tiny$\omega_{\mathrm{r}1}(k)$}
\psfrag{ddd}{\tiny$\omega_{\mathrm{r}1}\!(\!\mathrm{r}_{\mathrm{r}1}\!)$}
\psfrag{eee}{\tiny$\omega_{\mathrm{r}2}(1)$}
\psfrag{fff}{\tiny$\omega_{\mathrm{r}2}(2)$}
\psfrag{ggg}{\tiny$\omega_{\mathrm{r}2}(k)$}
\psfrag{hhh}{\tiny$\omega_{\mathrm{r}2}\!(\!\mathrm{r}_{\mathrm{r}2}\!)$}
\begin{center}
\subfloat[$P_\mathrm{r}^\mathrm{max} < P_\mathrm{ma}$, $\mu_{\rm
ma}^0 \geq \max\{\mu_1^0,
\mu_2^0\}$]{\label{f:wfB1}\includegraphics[width=0.42\textwidth]{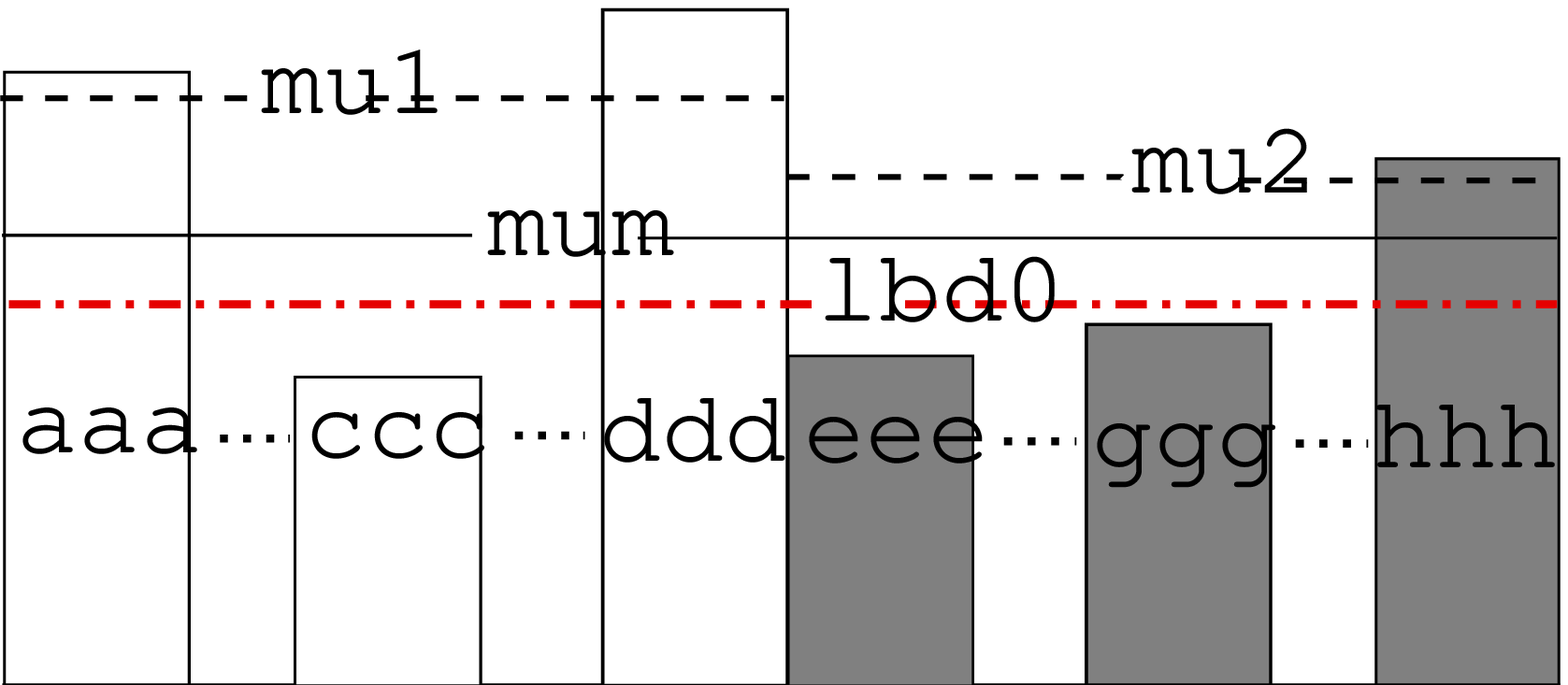}}
\hspace{2mm} \subfloat[$P_\mathrm{r}^\mathrm{max} <
\bar{P}_\mathrm{ma}$, $\mu_{\rm ma}^0 < \max\{\mu_1^0,
\mu_2^0\}$]{\label{f:wfB2}\includegraphics[width=0.42\textwidth]{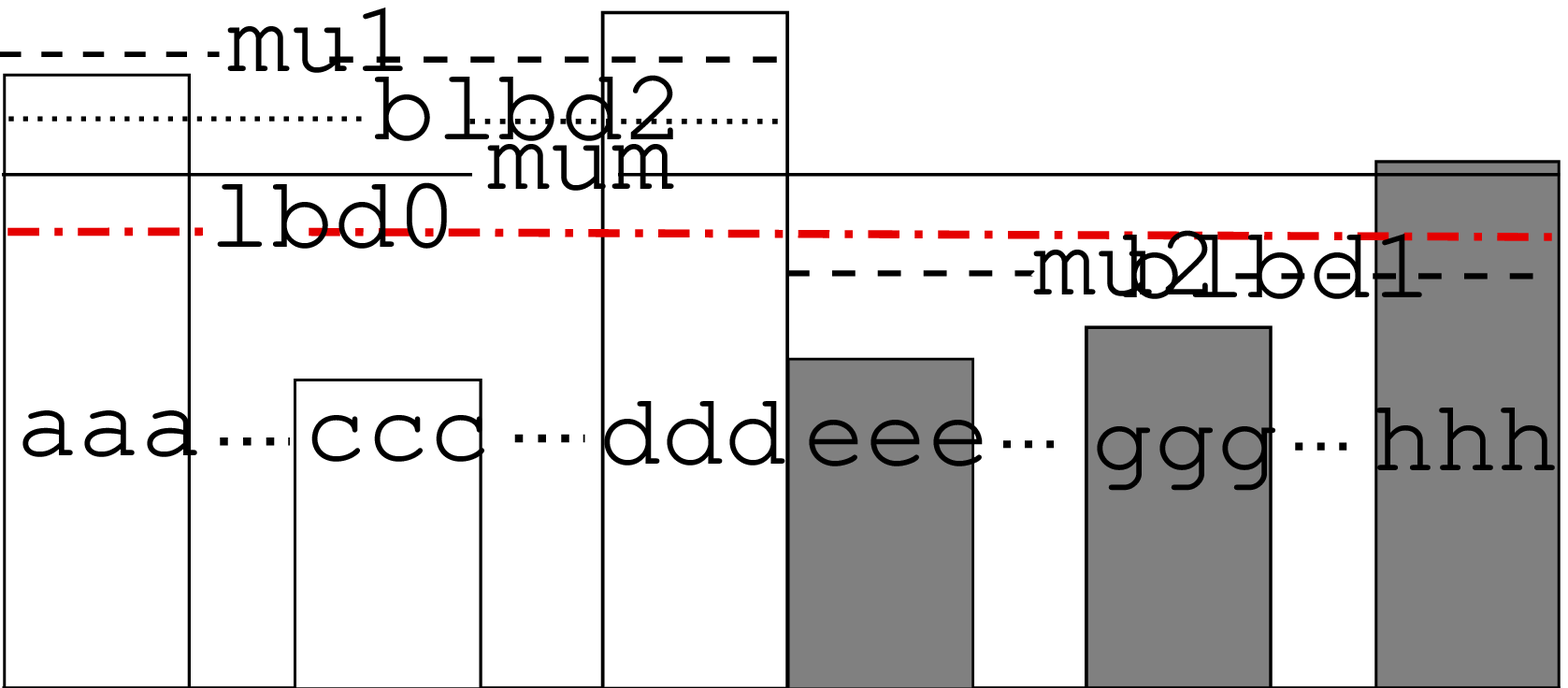}}
\end{center}
\caption {Illustration of $\mu_1^0$, $\mu_2^0$, $\mu_{\rm ma}^0$,
and $\lambda^0$ for the scenario of relay optimization.}
\label{f:wfB} \vspace{-0.2cm}
\end{figure}

Two of the above subcases, one for the subcase $P_\mathrm{r}^\mathrm{max} < P_\mathrm{ma}, \mu_{\rm
ma}^0 \geq \max\{\mu_1^0, \mu_2^0\}$ and the other for the subcase $P_\mathrm{r}^\mathrm{max} <
\bar{P}_\mathrm{ma}, \mu_{\rm ma}^0 < \max\{\mu_1^0, \mu_2^0\}$, are illustrated in Fig.~\ref{f:wfB}.

From the above discussion, it can be seen that the algorithm in
Table~\ref{t:RPA} obtains the optimal power allocation in at most
seven steps without iterations.

Recall that the sum-rate of DF TWR is bounded by both the sum-rate of the MA phase and the sum-rate of the BC phase. In the scenario of relay optimization, the relay optimizes its power allocation which affects the sum-rate of the BC phase. Since the relay may or may not use all its available power at optimality (i.e., for the optimal
power allocation), the sum-rate of the BC phase is not necessarily
maximized at optimality. Moreover, it is also possible
that the sum-rate of the BC phase at optimality is not
even the maximum sum-rate of the BC phase that can be achieved using the power
consumed by the relay at optimality. We specify the term
\emph{efficient} to describe such optimal power allocation of the relay
that maximizes the BC phase sum-rate $R^{\rm bc}(\mathbf{B})$
with the actually consumed power at the relay. Thus, the relay's power
allocation is efficient if it generates the maximum sum-rate for
broadcasting the messages of the source nodes given its power
consumption. For example, when the relay uses all its available power at optimality, the optimal power allocation of the relay is efficient if it maximizes the sum-rate of the BC phase, and inefficient otherwise. When the relay uses the power $P_\mathrm{r}<P_\mathrm{r}^{\max}$ at optimality, the optimal power allocation is efficient if the achieved sum-rate of the BC phase is the maximum achievable sum-rate of the BC phase with power consumption $P_\mathrm{r}$, and inefficient otherwise. Then the following two conclusions can be drawn for the scenario of
relay optimization.

First, the optimal relay power allocation in the scenario of relay
optimization is always efficient for the case that $\mu_{\rm ma}^0
\geq \max\{\mu_1^0, \mu_2^0\}$. \textcolor{Black}{In such a case, it can
be seen from \eqref{e:C1s1e1} and \eqref{e:C1s2e1} that $1/\lambda_1=1/\lambda_2$ at optimality regardless
of whether the relay uses all its available power. Therefore, the
BC phase sum-rate $R^{\rm bc}(\mathbf{B})$ is always maximized
given the relay's power consumption in this case.} However, the optimal relay power
allocation is inefficient for the case that $\mu_{\rm ma}^0 <
\max\{\mu_1^0, \mu_2^0\}$ as long as $P_{\rm r}^{\max}>P_{\rm l}$. Moreover, the larger the difference
between $\max\{\mu_1^0, \mu_2^0\}$ and $\mu_{\rm ma}^0$ in this case, the more
inefficient the optimal relay power allocation becomes when $P_{\rm r}^{\max}>P_{\rm l}$. Given
the definitions \eqref{e:watermu1}-\eqref{e:watermum} and Lemma~1,
the case with $\mu_{\rm ma}^0 < \max\{\mu_1^0, \mu_2^0\}$
indicates that one source node uses more power, has more
antennas and/or better channel condition compared to those of the
other source node. Indeed, if the power budget, number of
antennas, and channel conditions are the same for the two source
nodes, as an extreme example, it leads to $\mu_{\rm
ma}^0>\mu_1^0=\mu_2^0$. Therefore, it can be seen that the
asymmetry between the power budget, number of antennas, and/or
channel conditions can degrade the relay power allocation
efficiency in the scenario of relay optimization.

Second, the considered scenario of relay optimization may result in the
waste of power at the source nodes. 
However, the relay never wastes any power. This is due to the fact that the relay is aware of the
source node power allocation strategies and optimizes its own
power allocation based on them. As a result, it can use only part of the
available power if its power limit $P_\mathrm{r}^{\mathrm{max}}$
is large. However, the relay power
allocation strategy is unknown to the source nodes when the source nodes decide their power allocation strategies. Therefore, the possibility of wasting power in the relay optimization scenario
can be viewed as the tradeoff for low complexity.
Indeed, in the scenario of relay optimization, there is no
coordination between the relay and the source nodes. As a result,
it is almost impossible to achieve the maximum sum-rate with
minimum total power consumption referred to as network-level optimality. In order to achieve
the network-level optimality, the scenario of network
optimization, in which the relay and the source nodes jointly
maximize the sum-rate of the TWR with minimum power consumption,
is considered in Part II of this two-part paper.

\section{Simulations}\label{s:simula}
In this section, we provide simulation examples for some results
presented earlier and demonstrate the proposed algorithm for
relay optimization in Table~\ref{t:RPA}. The general setup is as
follows. The elements of the channels $\mathbf{H}_{\mathrm{r}i}$
and $\mathbf{H}_{i\mathrm{r}}, \forall i$ are generated from
complex Gaussian distribution with zero mean and unit covariance.
\textcolor{Black}{The noise variances $\sigma_i^2, \forall i$ and $\sigma_\mathrm{r}^2$ are equal to each other and denoted uniformly as $\sigma^2$}.
While the source node power allocation strategy $\mathbf{D}^{0}$ can be arbitrary, we use for simulations the $\mathbf{D}^{0}$ that maximizes the MA phase sum-rate $R^{\rm ma}(\mathbf{D})$. The rates $R^{\rm ma}(\mathbf{D})$, $\bar{R}_{i
\mathrm{r}} (\mathbf{D}_{i})$, and $\hat{R}_{\mathrm{r}i}
(\mathbf{B}_i )$ are briefly denoted as $R^{\rm ma}$,
$\bar{R}_{i\mathrm{r}}$ and $\hat{R}_{\mathrm{r}i}$, respectively,
in the figures in this section.

\begin{figure}[t]
\begin{center}
\includegraphics[angle=0,width=0.5\textwidth]{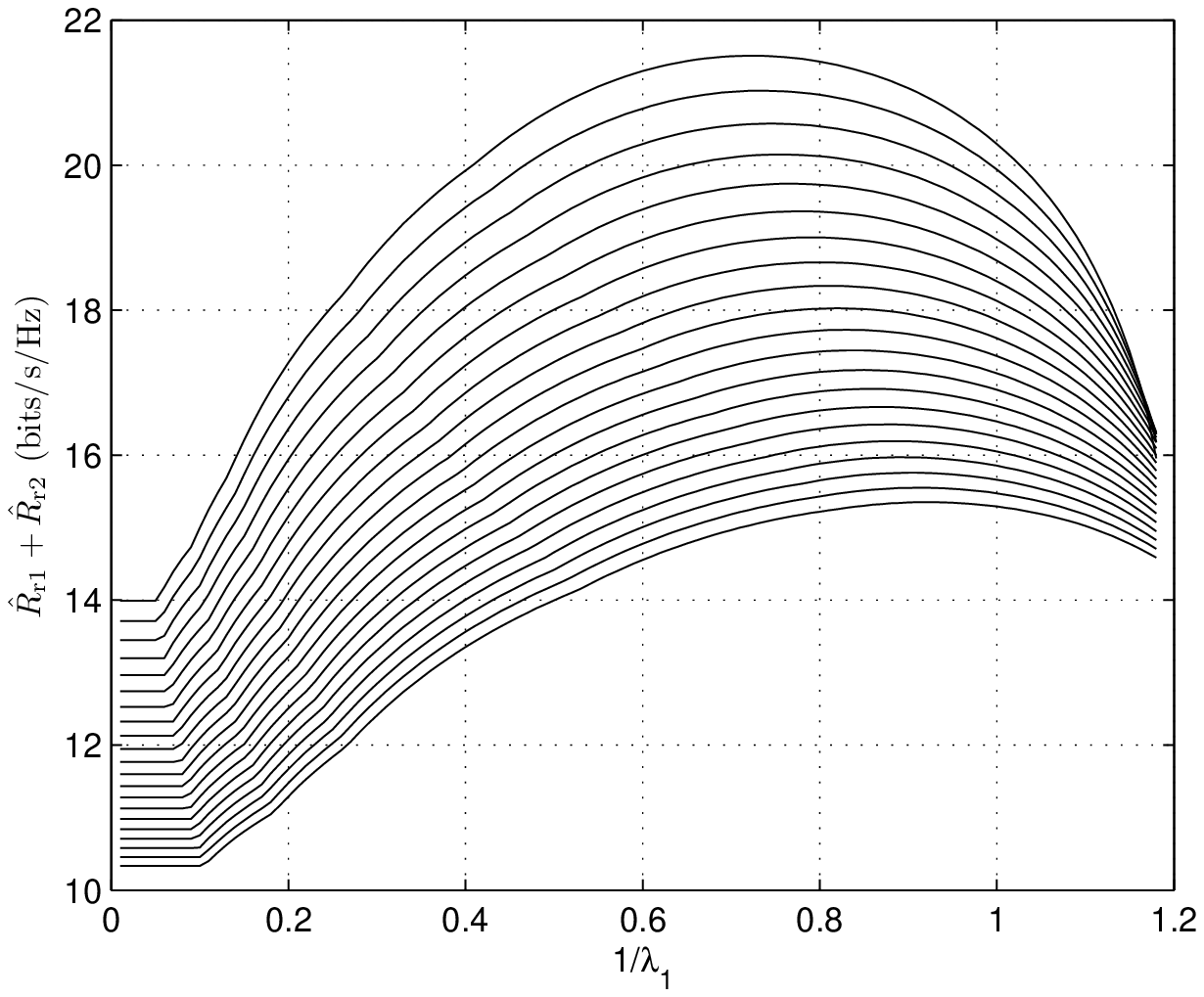}
\end{center}
\vspace{-4mm} \caption
{$\hat{R}_{\mathrm{r}1}+\hat{R}_{\mathrm{r}2}$ versus
$1/\lambda_1$ under different $P^\mathrm{max}_\mathrm{r}/\sigma^2$.} \label{f:lemma1}
\vspace{-5mm}
\end{figure}

\emph{Example 1: A demonstration of Lemma~2}. It is assumed that
the number of antennas at the relay $n_\mathrm{r}$ is 8 while
source node 1 has $n_1=6$ antennas and source node 2 has $n_2=5$
antennas. 
\textcolor{Black}{Each curve in Fig.~\ref{f:lemma1} shows the sum-rate
$\hat{R}_{\mathrm{r}1}+\hat{R}_{\mathrm{r}2}$ versus the
water-level $1/\lambda_1$ for a given ratio of $P^\mathrm{max}_\mathrm{r}$ over $\sigma^2$.
In each curve, for each given $1/\lambda_1$,
the relay consumes all the remaining power to maximize
$1/\lambda_2$. Therefore, the power consumption of the relay is
fixed and equals $P^\mathrm{max}_\mathrm{r}$. For each curve, $\sigma^2$ is different.
The curve at the bottom corresponds to the ratio $P^\mathrm{max}_\mathrm{r}/\sigma^2$ equal to $4~\text{dB}$. For each time, when the ratio of $P^\mathrm{max}_\mathrm{r}$ over $\sigma^2$ increases, a new curve
of $\hat{R}_{\mathrm{r}1}+\hat{R}_{\mathrm{r}2}$ versus
$1/\lambda_1$, which lies above the previous curve, is plotted. The curve at the top corresponds to the ratio $P^\mathrm{max}_\mathrm{r}/\sigma^2$ equal to $7~\text{dB}$.} 
It can be seen from Fig.~\ref{f:lemma1} that the sum-rate
$\hat{R}_{\mathrm{r}1}+\hat{R}_{\mathrm{r}2}$ is a nonconvex
function of $1/\lambda_1$. However,
$\hat{R}_{\mathrm{r}1}+\hat{R}_{\mathrm{r}2}$ is non-decreasing in
the interval from the minimum $1/\lambda_1$ to the sum-rate
maximizing $1/\lambda_1$ and non-increasing from the sum-rate
maximizing $1/\lambda_1$ to the maximum $1/\lambda_1$. \textcolor{Black}{
Note that $1/\lambda_1=1/\lambda_2=1/\lambda^0$ when the BC phase sum-rate is maximized.
As a result, it can be seen that increasing $\max\{1/\lambda_1, 1/\lambda_2\}$
and decreasing $\min\{1/\lambda_1, 1/\lambda_2\}$ while fixing the total power
consumption leads to a smaller BC phase sum-rate for any given
$\{1/\lambda_1, 1/\lambda_2\}$. Therefore, Fig.~\ref{f:lemma1}
verifies the result presented in Lemma~2.}

\begin{figure}[!t]
\centering \subfloat[$\hat{R}_{\mathrm{r}i}$ in the optimal
solution of the sum-rate maximizing problems with and without
minimizing power consumption, respectively, versus $P_{\rm r}^{\rm
max}$]
{\includegraphics[angle=0,width=0.47\textwidth]{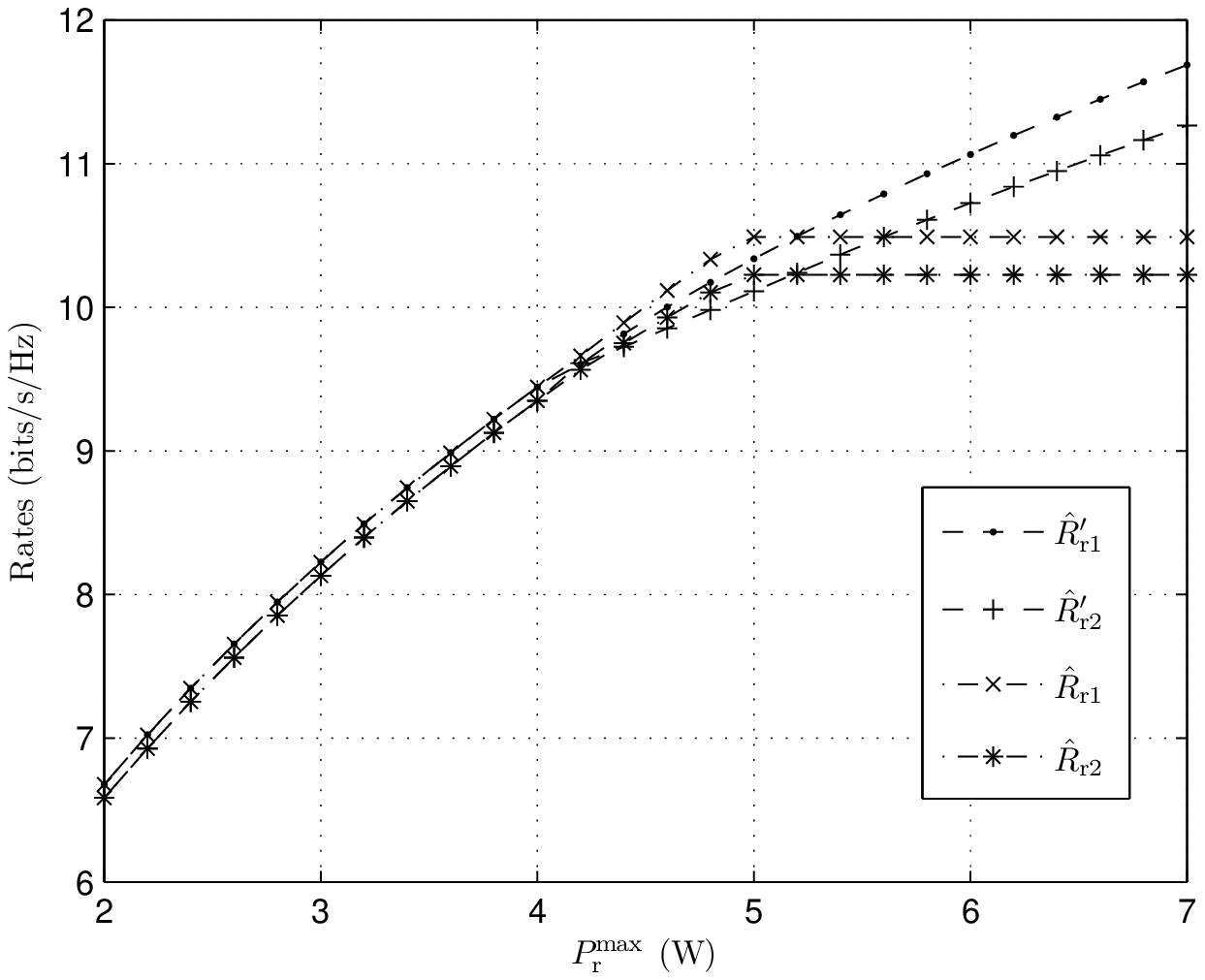}
\label{RelayOpt1}}
\hspace{5mm} \subfloat[Relay power consumption, $R^{\rm ma}(\mathbf{D}^0)$ and
$\sum\limits_i\hat{R}_{\mathrm{r}i}$ in the optimal solution of
the sum-rate maximizing problems with and without minimizing power
consumption, respectively, versus $P_{\rm r}^{\rm max}$]
{\includegraphics[angle=0,width=0.47\textwidth]{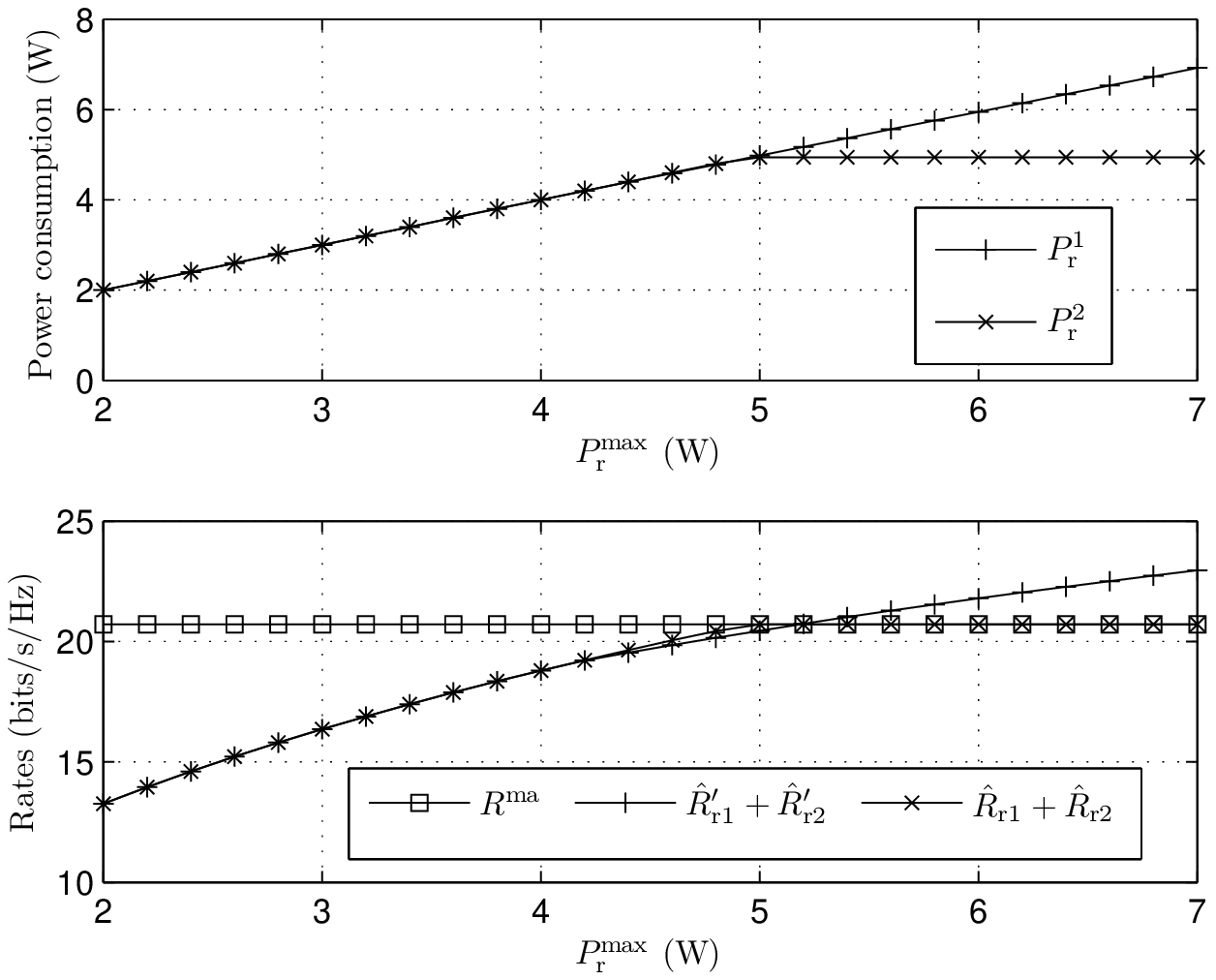}
\label{RelayOpt2}}
\caption{Illustration of relay optimization.}\label{RelayOpt}
\end{figure}

\emph{Example 2: The relay optimization problem.}
Fig.~\ref{RelayOpt1} compares the BC phase rates at optimality of the relay optimization problem, which considers
power consumption minimization, with the BC phase rates at optimality of the problem \eqref{e:Rl}, which does not minimize
the power consumption, under different $P_\mathrm{r}^{\mathrm{max}}$.
One channel realization is shown. The specific setup for this
simulation is as follows. The number of antennas $n_1, n_2$, and
$n_\mathrm{r}$ are set to be $6, 5$ and $8$, respectively. \textcolor{Black}{The power
limits for the source nodes are set to be $P_1^{\mathrm{max}} =
P_2^{\mathrm{max}} =3~\text{W}$. The noise variance is normalized so that $\sigma^2=1$}. The MA phase rates for this channel
realization are 20.7 for $R^{\mathrm{ma}} (\mathbf{D}^0)$, 11.2
for $\bar{R}_{1\mathrm{r}}(\mathbf{D}_1^0)$, and 11.0 for
$\bar{R}_{2\mathrm{r}}(\mathbf{D}_2^0)$. In Fig.~\ref{RelayOpt1},
$\hat{R}_{\mathrm{r}i}^{\prime}$ represents $\hat{R}_{\mathrm{r}i}
(\mathbf{B}_i^{\prime})$, where
$\mathbf{B}_i^{\prime}$'s, $\forall i$ are the optimal solution (\textcolor{Black}{obtained using CVX \cite{CVX}}) to the problem
\eqref{e:Rl} which does not minimize
the power consumption, and $\hat{R}_{\mathrm{r}i}$ represents
$\hat{R}_{\mathrm{r}i} (\mathbf{B}_i)$,  where $\mathbf{B}_i$'s, $\forall i$ are the optimal solution to the
relay optimization problem considering power consumption minimization obtained using the algorithm in
Table~\ref{t:RPA}. It can be seen from Fig.~\ref{RelayOpt1} that
$\hat{R}_{\mathrm{r}i}^{\prime} = \hat{R}_{\mathrm{r}i}$ when
$P_\mathrm{r}^{\mathrm{max}}$ is small. The reason is that
$\hat{R}_{\mathrm{r}i}^{\prime}$ is small when $P_\mathrm{r}^{\rm
max}$ is below certain threshold. As a result, the constraints in
$\eqref{e:RlconsPM}$ are always satisfied and the solutions to the
problem \eqref{e:Rl} and the relay optimization problem are the
same. As $P_\mathrm{r}^{\rm max}$ increases, $R^{\rm
tw}(\mathbf{B}, \mathbf{D}^0)$ becomes larger and is finally
bounded by $R^{\rm ma}(\mathbf{D}^0)$, while the relay
power consumption is not necessarily minimized in the solution of
the problem \eqref{e:Rl} which does not consider power
consumption minimization. This can be seen from the first subplot
of Fig.~\ref{RelayOpt2}, which shows that the power consumption in
the solution derived using the proposed algorithm, denoted as
$P_\mathrm{r}^2$, saturates when
$P_\mathrm{r}^{\mathrm{max}}\geq 4.9~\text{\textcolor{Black}{W}}$, while the power
consumption in the solution to the problem \eqref{e:Rl} which does not consider power
consumption minimization, denoted
as $P_\mathrm{r}^1$, keeps increasing. As a result, as can be seen
from the second subplot of Fig.~\ref{RelayOpt2},
$\sum\limits_i\hat{R}_{\mathrm{r}i}$ never exceeds
$R^{\mathrm{ma}}(\mathbf{D}^0)$, while $\sum\limits_i
\hat{R}_{\mathrm{r}i}^{\prime}$ grows beyond
$R^{\mathrm{ma}}(\mathbf{D}^0)$ when $R^{\rm tw}(\mathbf{B},
\mathbf{D}^0)$ is bounded by $R^{\mathrm{ma}}(\mathbf{D}^0)$.
\textcolor{Black}{Meanwhile, it can also be seen from the second subplot of Fig.~\ref{RelayOpt2}
that the maximum sum-rates $R^{\rm tw}(\mathbf{B}, \mathbf{D}^0)$ for the two compared solutions are
the same, both of which equal to $\sum\limits_i\hat{R}_{\mathrm{r}i}^{\prime} = \sum\limits_i\hat{R}_{\mathrm{r}i}$ when $\sum\limits\hat{R}_{\mathrm{r}i}^{\prime}\leq R^{\mathrm{ma}}(\mathbf{D}^0)$ and equal to $R^{\mathrm{ma}}(\mathbf{D}^0)$ when $\sum\limits\hat{R}_{\mathrm{r}i}^{\prime}> R^{\mathrm{ma}}(\mathbf{D}^0)$.} Thus, this example demonstrates
that the proposed algorithm in Table~\ref{t:RPA} achieves
maximum sum-rate in the scenario of relay optimization
with minimum power consumption.

\begin{figure}[!tt]\label{f:RelayAsym}
\centering \subfloat[Sum-rate at optimality.]
{\includegraphics[angle=0,width=0.47\textwidth]{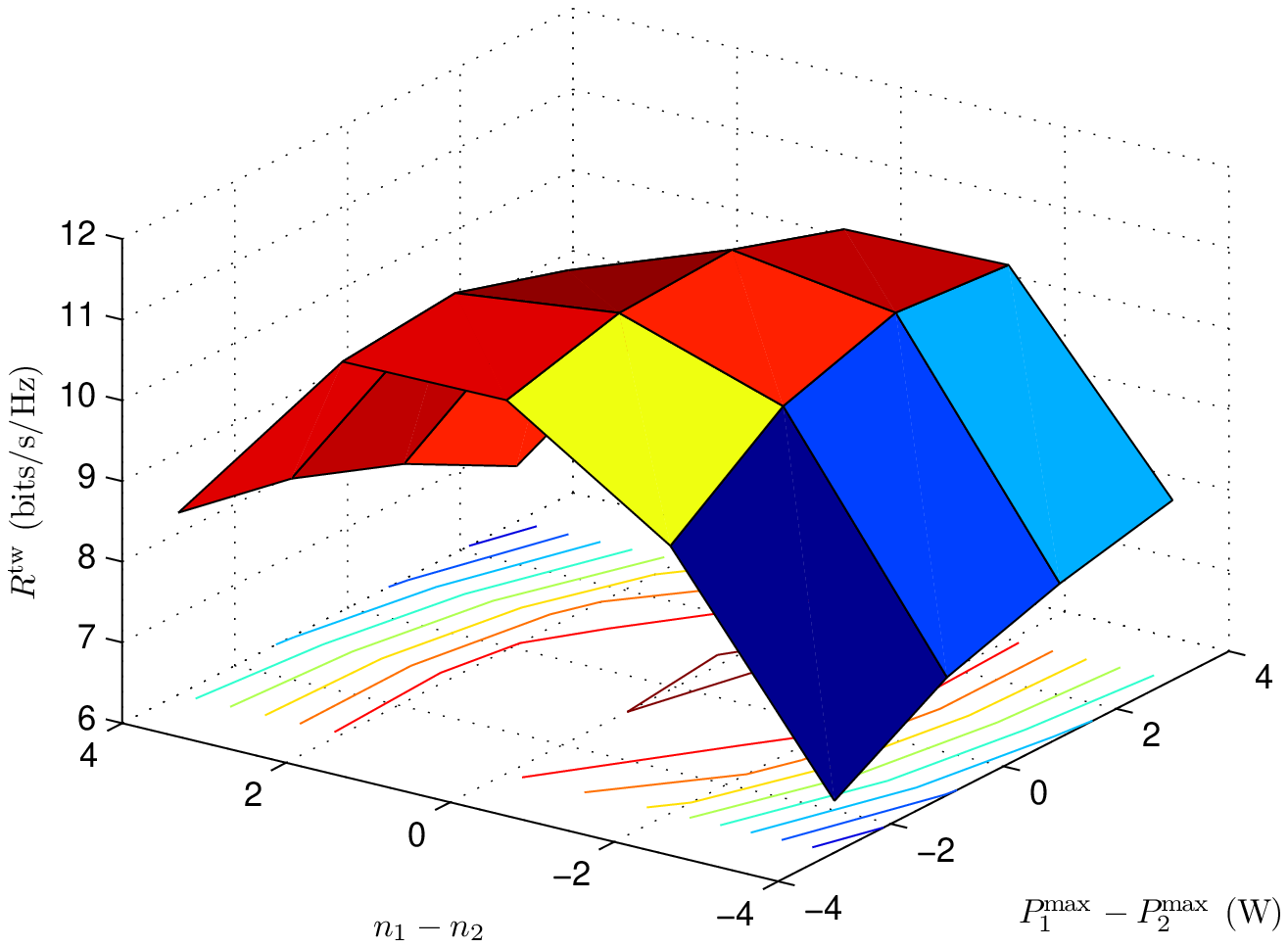}
\label{f:RelayAsymR}}
\hspace{5mm} \subfloat[Relay power consumption at optimality.]
{\includegraphics[angle=0,width=0.47\textwidth]{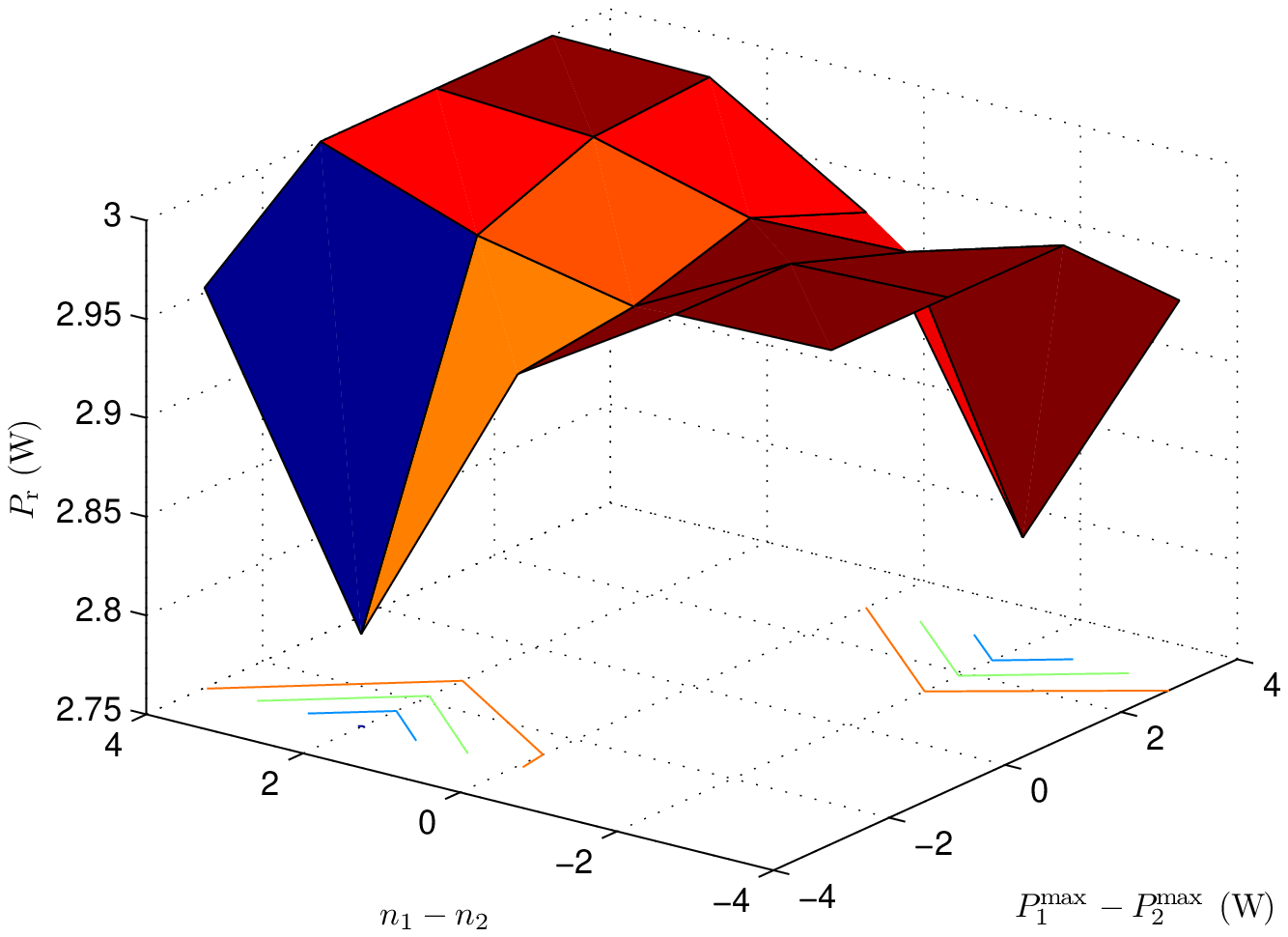}
\label{f:RelayAsymP}}
\hspace{5mm} \subfloat[Percentage of efficient power allocation at optimality.]
{\includegraphics[angle=0,width=0.47\textwidth]{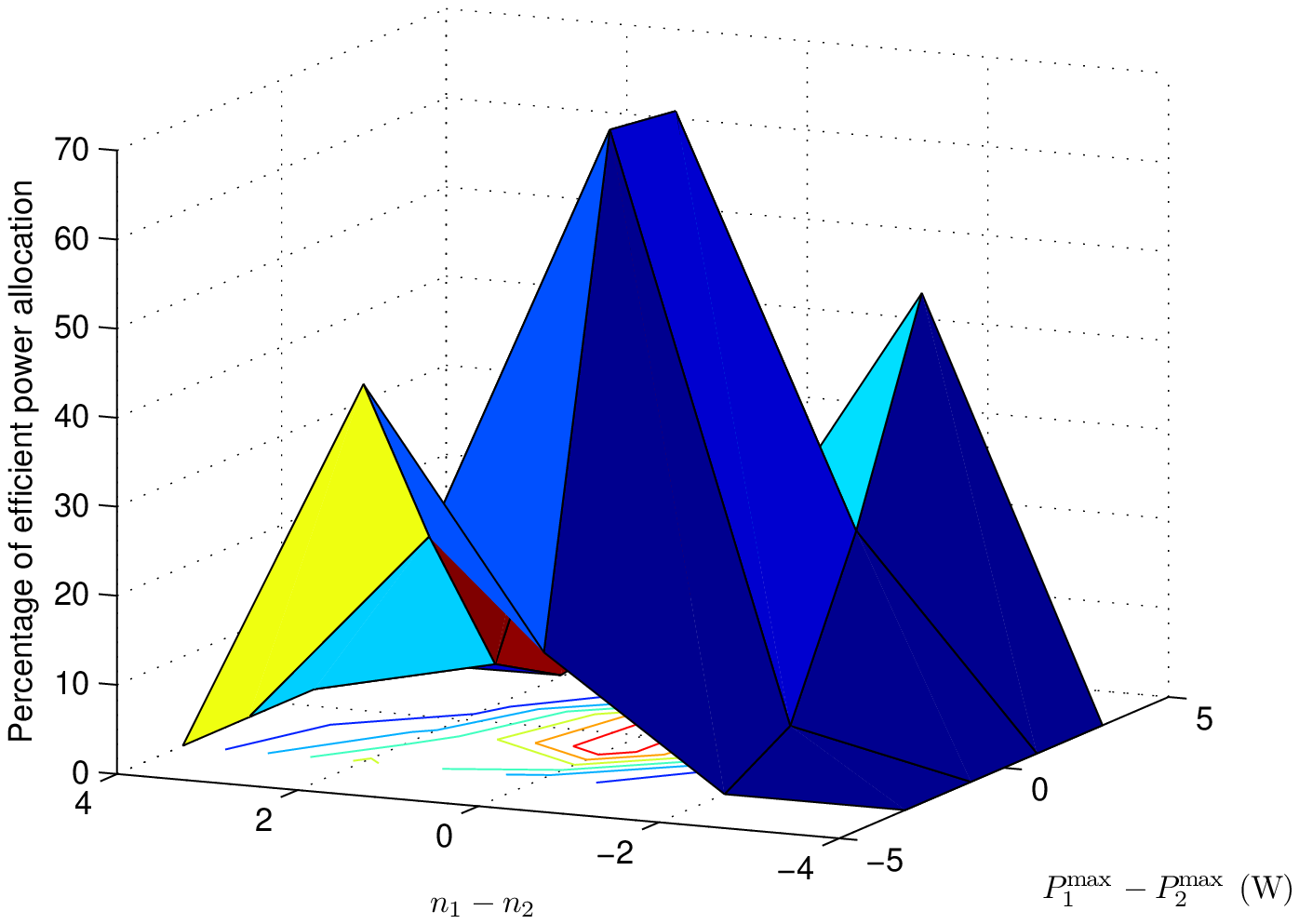}
\label{f:RelayAsymE}}
\caption{Effect of asymmetry: the average sum-rate, average relay power consumption, and percentage of efficient power allocation at optimality of relay optimization versus the difference between number of antennas and the difference between power limits at the source nodes in 1000 channel realizations.}\label{f:RelayAsym}
\end{figure}

\emph{Example 3: The effect of asymmetry}. The specific setup for this example is as
follows. The noise variance is normalized so that $\sigma^2=1$. The number of antennas at the relay, i.e., $n_\mathrm{r}$, is set to be $6$. The power limit of the relay, i.e., $P^{\rm max}_\mathrm{r}$ is set to be 3~\text{\textcolor{Black}{W}}. The total number of antennas at both source nodes is fixed such that $n_1+n_2=6$. The total available power at both source nodes is also fixed such that $P^{\rm max}_1+P^{\rm max}_2=5~\text{\textcolor{Black}{W}}$. Given the above total number of antennas and total available power at the source nodes, the relay optimization problem is solved for different $n_1$, $n_2$, $P^{\rm max}_1$, and $P^{\rm max}_2$ for 1000 channel realizations. The resulting average sum-rate and average power consumption of the relay, and the percentage of efficient power allocation at optimality are plotted in Figs.~\ref{f:RelayAsymR},~\ref{f:RelayAsymP}~and~\ref{f:RelayAsymE}, respectively, versus the difference between the number of antennas and the difference between the power limits at the source nodes. From Fig.~\ref{f:RelayAsymR}, it can be seen that the sum-rate at optimality of the relay optimization is the largest when there is no asymmetry in the number of antennas at the source nodes and no asymmetry or only small asymmetry in the power limits of the source nodes. As the asymmetry becomes larger in either number of antennas or power limits, the sum-rate at optimality of the relay optimization decreases. Therefore, it can be seen from this figure that the asymmetry in the above aspects leads to smaller sum-rate at optimality of the considered relay optimization problem. Relating Figs.~\ref{f:RelayAsymP}~and~\ref{f:RelayAsymE} to Fig.~\ref{f:RelayAsymR}, two more observations can be made. First, the relay does not necessarily use all the available power for sum-rate maximization in the relay optimization scenario. Second, the asymmetry in number of antennas and power limits leads to low power allocation efficiency. It can be seen from Fig.~\ref{f:RelayAsymP} that when one of $P_1^{\rm max}-P_2^{\rm max}$ and $n_1-n_2$ is positive while the other is negative, the relay uses a part of its available power. However, the achieved sum-rate is smaller compared to the sum-rate in the case when $P_1^{\rm max}-P_2^{\rm max}=0$ and $n_1-n_2=0$ (see Fig.~\ref{f:RelayAsymR}). In this situation, since the average power consumption and the average sum-rate are both low, the percentage of efficient power allocation is larger than 0 but less than the percentage when $P_1^{\rm max}-P_2^{\rm max}=0$ and $n_1-n_2=0$, as can be seen from Fig.~\ref{f:RelayAsymE}. When $P_1^{\rm max}-P_2^{\rm max}$ and $n_1-n_2$ are  both positive or  both negative, the relay uses more power than the power used in the case when $P_1^{\rm max}-P_2^{\rm max}=0$ and $n_1-n_2=0$ while the achieved sum-rate is smaller than that in the latter case. In this situation, since the average power consumption is high while the average sum-rate is low, the percentage of efficient power allocation is very low, if not zero, as can be seen from Fig.~\ref{f:RelayAsymE}. The above facts become more obvious when the asymmetry becomes larger. Therefore, it can be seen from Figs.~\ref{f:RelayAsymP}~and~\ref{f:RelayAsymE} that the asymmetry on the power limits
and the number of antennas can lead to low power allocation efficiency.

\section{Conclusion}\label{s:conclu}
In Part I of this two-part paper, we have solved the problem of sum-rate maximization with minimum power consumption for MIMO DF TWR in the scenario of relay optimization. For finding the optimal solution, we have proved the sufficient and necessary optimality condition for power allocation. Based on this condition, we have proposed an algorithm to find the optimal solution. The proposed algorithm allows the relay to obtain its optimal power allocation in several steps. We have shown that, as a trade-off for low complexity, there can be waste of power at the source nodes in the relay optimization scenario because of the lack of coordination.  We have also shown that the asymmetry in the number of antennas and power limits at the source nodes can result in the degradation of the sum-rate performance and the power allocation efficiency in MIMO DF TWR. \textcolor{Black}{Next, in Part~II of this two-part paper, we will investigate the scenario in which the relay and the source nodes jointly optimize their transmit strategies to achieve the network-level optimality of sum-rate maximization with minimum total power consumption for the MIMO DF TWR.}

\section{Appendix}\label{s:appen}

\subsection{Proof of Lemma~2}\label{s:PLm2}
Lemma~2 is proved in two steps, \textcolor{Black}{i.e., Steps~A~and~B.
In Step~A, we prove that $\sum\limits_l \hat{R}_{\mathrm{r}l}(\lambda_l^{\prime})$
can be increased by modifying the current power allocation on two specific subchannels.
In Step~B, we show that $\sum\limits_l \hat{R}_{\mathrm{r}l}(\lambda_l^{\prime})$
may be further increased.}

\textcolor{Black}{Step~A: $\sum\limits_l \hat{R}_{\mathrm{r}l}(\lambda_l^{\prime})$
can be increased.} Given the fact that $\sum\limits_{l} \text{Tr} \{
\mathbf{P}_{\mathrm{r}l}(\lambda_l)\} =\sum\limits_{l} \text{Tr}
\{ \mathbf{P}_{\mathrm{r}l} (\lambda_l^{\prime})\}$, it can be
shown that $1/\lambda_i^{\prime} > \min\limits_k\{ 1/ \alpha_i
(k)\}$ as long as $1/\lambda_j > \min\limits_k\{1 / \alpha_j
(k)\}$. As a result, there exist $k_1$ and $k_2$ such that
$1/\lambda_i^{\prime} > 1/ \alpha_i (k_1)$ and $1/ \lambda_j >
1/\alpha_j(k_2)$. Define $f(p_{\mathrm{r}i}(k_1)) = \log \big( 1 +
\alpha_i(k_1) p_{\mathrm{r}i} (k_1) \big) + \log \big( 1 +
\alpha_j (k_2) p_{\mathrm{r}j}(k_2)\big)$ where $p_{\mathrm{r}j}
(k_2)=p-p_{\mathrm{r}i} (k_1)$ and $p$ is a positive constant. It
can be seen that $f(p_{\mathrm{r}i}(k_1))$ is strictly concave in
$p_{\mathrm{r}i}(k_1)\in[0, p], \forall p>0$. Set $p = \big(
1/\lambda_j^{'} - 1 / \alpha_j(k_2) \big)^{+} +1/
\lambda_i^{\prime} - 1/\alpha_i(k_1)$. The optimal allocation of
the power $p$ on $\alpha_i(k_1)$ and $\alpha_j(k_2)$ that
maximizes $f(p_{\mathrm{r}i} (k_1))$ is $p_{\mathrm{r}i}
(k_1)=\big(1/\lambda^{\rm opt}(p)-1/\alpha_i(k_1)\big)^{+}$ and
$p_{\mathrm{r}j}(k_2)=\big(1/\lambda^{\rm opt}(p)-1 /
\alpha_j(k_2) \big)^{+}$ where $\lambda^{\rm opt}(p)$ is a
function of $p$ and $1/\lambda^{\rm opt}(p)$ is the optimal water
level. It can be shown that $1/\lambda^{\rm opt}(p)
<1/\lambda_i^{\prime}$. There exist two cases, i.e., $1/
\lambda^{\rm opt} (p) \leq 1/\lambda_i$ and $1/\lambda^{\rm
opt}(p) > 1/\lambda_i$. In the case when $1 /\lambda^{\rm opt}(p)
\leq 1/ \lambda_i$, it follows that $\big(1/\lambda^{\rm
opt}(p)-1/\alpha_i(k_1)\big)^{+}\leq \big( 1/ \lambda_i-1 /
\alpha_i(k_1))^{+} < 1/ \lambda_i^{\prime}-1/ \alpha_i(k_1)$.
The power allocation on $k_1$ and $k_2$ using $\lambda_i^{\prime}$
and $\lambda_j^{\prime}$ is
\begin{subequations}\label{e:cpak12}
\vspace{-2mm}
\begin{align}
&p_{\mathrm{r}i}(k_1)=\bigg(\frac{1}{\lambda_i^{\prime}} - \frac{1}
{\alpha_i(k_1)}\bigg)^{+} \label{e:cpak1}\\
&p_{\mathrm{r}j}(k_2)=\bigg(\frac{1}
{\lambda_j^{\prime}}-\frac{1}{\alpha_j(k_2)}\bigg)^{+}.\label{e:cpak2}
\end{align}
\end{subequations}
Since $f(p_{\mathrm{r}i}(k_1))$ is strictly concave as mentioned
above, it can be seen that the power allocation
\begin{subequations}\label{e:leprijk12}
\vspace{-2mm}
\begin{align}
p_{\mathrm{r}i}(k_1)&=\bigg(\frac{1}{\lambda_i}-\frac{1}
{\alpha_i(k_1)}\bigg)^{+} \\
p_{\mathrm{r}j}(k_2)&=\bigg(\frac{1}{\lambda_j^{\prime}}
-\frac{1}{\alpha_j(k_2)}\bigg)^{+} \nonumber\\
&\;\;\;+\frac{1}{\lambda_i^{\prime}}-\frac{1}{\alpha_i(k_1)} -
\bigg(\frac{1}{\lambda_i}-\frac{1}{\alpha_i(k_1)}
\bigg)^{+}
\end{align}
\end{subequations}
which reduces $p_{\mathrm{r}i}(k_1)$ and increases
$p_{\mathrm{r}j}(k_2)$, both by ${1} / {\lambda_i^{\prime}}-{1} /
{\alpha_i(k_1)}-\big({1}/{\lambda_i}-{1}/{\alpha_i(k_1)}\big)^{+}$,
yields higher $f(p_{\mathrm{r}i}(k_1))$ than the power allocation
in \eqref{e:cpak12}.

Therefore, the sum-rate $\sum\limits_l\sum\limits_k \log \big(
1+\alpha_l(k)p_{\mathrm{r}l}(k)\big)$ achieved using
\eqref{e:leprijk12} and
\begin{subequations}\label{e:leprijk}
\vspace{-2mm}
\begin{align}
&p_{\mathrm{r}i}(k)=\bigg(\frac{1}{\lambda_i^{\prime}} -
\frac{1}{\alpha_i(k)}\bigg)^{+}, \forall k\in \mathcal{I}_i
\setminus\{k_1\} \\
&p_{\mathrm{r}j}(k)=\bigg(\frac{1}{\lambda_j^{\prime}} -
\frac{1}{\alpha_j(k)}\bigg)^{+}, \forall k\in \mathcal{I}_j
\setminus\{k_2\}
\end{align}
\end{subequations}
is larger than $\sum\limits_l \hat{R}_{\mathrm{r}l}
(\lambda_l^{\prime})$. This is the first step of increasing
sum-rate. Moreover, it can be seen that there exists
$\tilde{\lambda}_j$ such that
\begin{subequations}
\vspace{-1mm}
\begin{align}
&\qquad\qquad\qquad\qquad\frac{1}{\lambda_j^{\prime}} < \frac{1}{\tilde{\lambda}_j} <
\frac{1}{\lambda_j} \\
&\text{Tr}\{\mathbf{P}_{\mathrm{r}i}(\lambda_i^{\prime})\} -
\bigg(\frac{1}{\lambda_i^{\prime}}-\frac{1}{\alpha_i(k_1)}
\bigg)^{+} +\bigg( \frac{1}{\lambda_i} - \frac{1}{\alpha_i(k_1)}
\bigg)^{+} \nonumber \\
&\qquad\qquad+ \text{Tr} \{\mathbf{P}_{\mathrm{r}j}
(\tilde{\lambda}_j) \} = \sum\limits_{l} \text{Tr} \{
\mathbf{P}_{\mathrm{r}l} (\lambda_l^{\prime})\}
\end{align}
\vspace{-1mm}
\end{subequations}
and the power allocation
\begin{subequations}\label{e:leprijknew}
\vspace{-1mm}
\begin{align}
&p_{\mathrm{r}i}(k_1)=\bigg(\frac{1}{\lambda_i} -
\frac{1}{\alpha_i(k)}\bigg)^{+} \\
&p_{\mathrm{r}i}(k) =
\bigg(\frac{1}{\lambda_i^{\prime}} - \frac{1}{\alpha_i
(k)}\bigg)^{+}, \forall k \in \mathcal{I}_i\setminus\{k_1\}\\
&p_{\mathrm{r}j}(k)=\bigg(\frac{1}{\tilde{\lambda}_j} -
\frac{1}{\alpha_j(k)}\bigg)^{+}, \forall k\in \mathcal{I}_j
\end{align}
\vspace{-1mm}
\end{subequations}
which spreads the power ${1}/{\lambda_i^{\prime}}-{1} /
{\alpha_i(k_1)}-\big({1}/{\lambda_i}-{1}/{\alpha_i(k_1)}\big)^{+}$
over ${\alpha_j(k)}$'s, $\forall k\in \mathcal{I}_j$, achieves even higher
sum-rate than that achieved by the power allocation specified by
\eqref{e:leprijk12} and \eqref{e:leprijk}. This is the second step
of increasing the sum-rate.

For the second case in which $1/\lambda_i <1/\lambda^{\rm
opt}(p)<1/\lambda_i^{\prime}$, the following process is adopted.
Similar to the two steps of increasing the sum-rate in the first
case, the sum-rate $\sum\limits_l\sum\limits_k
\log\big(1+\alpha_l(k)p_{\mathrm{r}l}(k)\big)$ increases after
each of the following two adjustments of power allocation.
First, reduce $p_{\mathrm{r}i}(k_1)$ from $1 / \lambda_i^{\prime}
- 1/\alpha_i(k_1)$ to $\big(1/\lambda^{\rm opt}(p) -
1/\alpha_i(k_1) \big)^{+}$. Then, spread the reduced power
${1}/{\lambda_i^{\prime}}-{1}/{\alpha_i(k_1)}-\big({1}/{\lambda^{\rm
opt}(p)}-{1}/{\alpha_i(k_1)}\big)^{+}$ over ${\alpha_j(k)}$'s $, k\in
\mathcal{I}_j$ by finding and using $1/\tilde{\lambda}_j^{\prime}$
which satisfies
\begin{eqnarray}
&\text{Tr}\{\mathbf{P}_{\mathrm{r}i}(\lambda_i^{\prime})\} -
\bigg(\frac{1}{\lambda_i^{\prime}}-\frac{1}{\alpha_i(k_1)}\bigg)^{+}
+\bigg(\frac{1}{\lambda^{\rm opt}(p)}-\frac{1}{\alpha_i(k_1)}
\bigg)^{+} \nonumber\\
&+ \text{Tr} \{ \mathbf{P}_{\mathrm{r}j}
(\tilde{\lambda}_j^{\prime})\} = \sum\limits_{l} \text{Tr} \{
\mathbf{P}_{\mathrm{r}l}(\lambda_l^{\prime})\}.
\end{eqnarray}
After the adjustments, it is
straightforward to see that the total power allocated on $k_1$ and
$k_2$ is reduced from $p = \big(1/\lambda_j^{\prime}-1/
\alpha_j(k_2)\big)^{+}+1/\lambda_i^{\prime}-1/\alpha_i(k_1)$ to
$\bar{p}=\big(1/\tilde{\lambda}_j^{\prime}-1/\alpha_j(k_2)\big)^{+}
+\big(1/\lambda^{\rm opt}(p)-1/\alpha_i(k_1)\big)^{+}$. In
consequence, there exists a new optimal water level $1/
\lambda^{\rm opt}(\bar{p})$ based on which the optimal allocation
of the power $\bar{p}$, i.e., $p_{\mathrm{r}i}(k_1) = \big(1/
\lambda^{\rm opt}(\bar{p})-1/\alpha_i(k_1)\big)^{+}$ and
$p_{\mathrm{r}j}(k_2)=1/\lambda^{\rm opt} (\bar{p}) - 1/
\alpha_j(k_2)$, maximizes $f(p_{\mathrm{r}i}(k_1))$ when $p$ in
$f(p_{\mathrm{r}i}(k_1))$ is substituted by $\bar{p}$. Since
$\bar{p}<p$, it can be seen that $1/\lambda^{\rm opt}
(\bar{p})<1/\lambda^{\rm opt}(p)$. Update $p$ and $1/\lambda^{\rm
opt}(p)$ so that $p=\bar{p}$ and $1/ \lambda^{\rm opt} (p) = 1 /
\lambda^{\rm opt} (\bar{p})$. Then the above process of reducing
$p_{\mathrm{r}i}(k_1)$ to $\big(1/\lambda^{\rm opt} (p)- 1 /
\alpha_i(k_1) \big)^{+}$, finding the new $1 /
\tilde{\lambda}_j^{\prime}$ and the new $1/\lambda^{\rm opt}(p)$
can be repeated until a). $1/\lambda^{\rm opt}(p)\leq 1 / \lambda_i$
or until b). $1/\lambda^{\rm opt}(p)\leq 1 / \alpha_i(k_1)$. The
former matches the condition for the first case discussed
in the previous paragraph and therefore can
be dealt with in the same way as in the first case, which leads to
\eqref{e:leprijknew}. The latter implies that $1 / \lambda_i < 1/
\lambda^{\rm opt}(p)\leq 1/\alpha_i(k_1)$, in which case the power
allocation can also be equivalently written as
\eqref{e:leprijknew}. Note that during this process the sum-rate
$\sum\limits_l\sum\limits_k \log\big(1 + \alpha_l(k) p_{\mathrm{r}
l} (k) \big)$ increases. Therefore, summarizing the above two
cases of $1/ \lambda^{\rm opt}(p) \leq 1/\lambda_i$ and
$1/\lambda^{\rm opt}(p) > 1/\lambda_i$, it is proved that the
sum-rate can be increased by reducing $p_{\mathrm{r}i}(k_1)$ from
$1/\lambda_i^{\prime}-1/\alpha_i(k_1)$ to $\big(1/ \lambda_i-1 /
\alpha_i(k_1) \big)^{+}$ and using the power allocation in
\eqref{e:leprijknew}.

\textcolor{Black}{Step~B: $\sum\limits_l \hat{R}_{\mathrm{r}l}(\lambda_l^{\prime})$
may be further increased.} Keep the above selected $k_2$ unchanged. As long as there
exists $k$ such that $p_{\mathrm{r}i}(k)= \big(1/
\lambda_i^{\prime}- 1/\alpha_i(k_1)\big)^{+}$ and
$p_{\mathrm{r}i}(k)>0$, this $k$ can be selected as $k_1$ and the
procedure of reducing $p_{\mathrm{r}i}(k_1)$ from
$1/\lambda_i^{\prime}-1/\alpha_i(k_1)$ to
$\big(1/\lambda_i-1/\alpha_i(k_1)\big)^{+}$ and spreading the
reduced power over ${\alpha_j(k)}$'s, $\forall k\in \mathcal{I}_j$ as
specified in \eqref{e:leprijknew} can be performed. This process
can be repeated until $p_{\mathrm{r}i}(k) = \big(1/\lambda_i -
1/\alpha_i(k)\big)^{+}, \forall k \in \{q\in\mathcal{I}_i|\big(
1/\lambda_i^{\prime} -1/\alpha_i(q)\big)^{+}>0\}$ and
$p_{\mathrm{r}i}(k) =0, \forall k \in \{q\in \mathcal{I}_i|\big(1/
\lambda_i^{\prime} -1/ \alpha_i(q)\big)^{+} = 0\}$. Note that the
sum-rate $\sum\limits_l\sum\limits_k \log\big(1+\alpha_l(k)
p_{\mathrm{r}l} (k)\big)$ increases in the above process for every
qualifying $k_1$. The resulting power allocation on $\alpha_i(k)$'s, $\forall
k\in \mathcal{I}_i$ is equivalent to $p_{\mathrm{r}i}(k)=\big(1/
\lambda_i-1/\alpha_i(k)\big)^{+}, \forall k\in \mathcal{I}_i$
since $\big(1/\lambda_i-1/\alpha_i(k) \big)^{+}= 0$ if
$\big(1/\lambda_i^{\prime}-1/\alpha_i(k) \big)^{+}= 0$. From the
procedure described in the previous paragraphes, the resulting power
allocation on $\alpha_j(k)$'s, $\forall k\in \mathcal{I}_j$ is
$p_{\mathrm{r}j}(k) = \big(1/\tilde{\lambda}_j-1 /\alpha_j(k)
\big)^{+}, \forall k$. According to the power constraint
$\sum\limits_{l} \text{Tr} \{ \mathbf{P}_{\mathrm{r}l} (\lambda_l)
\} =\sum\limits_{l} \text{Tr}\{ \mathbf{P}_{\mathrm{r}l}
(\lambda_l^{\prime})\}$ and the fact that the total power
consumption is fixed at all time, it can be seen that
$1/\tilde{\lambda}_j = 1/\lambda_j$.

Summarizing the above two steps, Lemma~2 is proved.
\hfill$\blacksquare$

\subsection{Proof of Lemma~3}\label{s:PLm3}

Given that $\lambda_i^{\prime}\leq\lambda_j$, we have $\lambda_i <
\lambda_i^{\prime}\leq\lambda_j<\lambda_j^{\prime}$. According to
Lemma~2, there exists $\tilde{\lambda}_i<\lambda_i^{\prime}$ such
that
\begin{eqnarray}
&\text{Tr}\{\mathbf{P}_{\mathrm{r}i}(\lambda_i^{\prime})\}+\text{Tr}
\{\mathbf{P}_{\mathrm{r}j}(\lambda_j)\} \qquad\qquad \nonumber \\
&\qquad\quad\quad\;\,=\text{Tr}\{\mathbf{P}_{
\mathrm{r}i}(\tilde{\lambda}_i)\}+\text{Tr}\{
\mathbf{P}_{\mathrm{r}j}(\lambda_j^{\prime})\}
\end{eqnarray}
and
\begin{eqnarray}
\hat{R}_{\mathrm{r}i}(\lambda_{i}^{\prime})+ \hat{R}_{\mathrm{r}j}
(\lambda_{j}) > \hat{R}_{\mathrm{r}i}( \tilde{\lambda}_i)+
\hat{R}_{\mathrm{r}j} (\lambda_{j}^{\prime}).
\end{eqnarray}

Therefore, given that
\begin{eqnarray}
\hat{R}_{\mathrm{r}i} (\lambda_{i}^{\prime})+
\hat{R}_{\mathrm{r}j} (\lambda_{j}) = \hat{R}_{\mathrm{r}i}
(\lambda_i)+ \hat{R}_{\mathrm{r}j} (\lambda_{j}^{\prime})
\end{eqnarray}
it is necessary that $\tilde{\lambda}_i>\lambda_i$. As a result,
it leads to
\begin{eqnarray}
&\text{Tr}\{\mathbf{P}_{\mathrm{r}i}(\lambda_i^{\prime})\}+\text{Tr}
\{\mathbf{P}_{\mathrm{r}j}(\lambda_j)\} \qquad\qquad\nonumber \\
&\qquad\quad\;\,<\text{Tr}\{\mathbf{P}_{
\mathrm{r}i}(\lambda_i)\}+ \text{Tr}\{
\mathbf{P}_{\mathrm{r}j}(\lambda_j^{\prime})\}.
\end{eqnarray}
Lemma~3 is thereby proved. \hfill$\blacksquare$

\subsection{Proof of Theorem~1}\label{s:PTh1}

First we prove that the optimal water-levels must satisfy condition \eqref{e:optpropty1}. It can be seen that the maximum $R^{\rm tw}(\mathbf{B}, \mathbf{D})$ is achieved with minimum power consumption using $\lambda_1 = \lambda_2 = \max \{\lambda^0, \mu_\mathrm{ma}^0\}$ when $\min\{1/\mu_1^0, 1/\mu_2^0\}\geq 1/\mu_\mathrm{ma}$ at the
optimality. Therefore, it is necessary that $\min\{1/\mu_1^0, 1/\mu_2^0\}< 1/\mu_{\rm ma}^0$ given that $\lambda_1 \neq \lambda_2$ at optimality. Let us consider the case when $\min\{1/\lambda_1, 1/\lambda_2\}=1/\lambda_1 < 1/\lambda_2$ at optimality. According to the constraint \eqref{e:WLcons1}, we have that $1/\lambda_1\leq 1/\mu_2^0$ at optimality. Similarly, it can be seen that $1/\lambda_2\leq 1/\mu_1^0$ at optimality. Since $1/\lambda_1 < 1/\lambda_2$, it leads to the result that $1/\lambda_1\leq 1/\mu_2^0<1/\mu_1^0$ at optimality. Assuming that $\min\{1/\mu_1^0, 1/\mu_2^0\}\neq 1/\lambda_1$ at optimality when $\lambda_1 \neq \lambda_2$, it infers that $1/\lambda_1<1/\mu_2^0<1/\lambda_2$. However, it can be seen that the power allocation using $1/\lambda_1<1/\mu_2^0<1/\lambda_2$ does not provide the maximum achievable $R^{\rm tw}(\mathbf{B}, \mathbf{D})$ according to Lemma~2. Consequently, the resulting power allocation is not optimal. It contradicts the assumption that $\min\{1/\mu_1^0, 1/\mu_2^0\}\neq 1/\lambda_1$ at optimality. Thus, the above assumption is invalid and it is necessary that $\min\{1/\mu_1^0, 1/\mu_2^0\}= 1/\lambda_1$ at optimality when $\lambda_1 \neq \lambda_2$. Similarly, it can be proved that $\min\{1/\mu_1^0, 1/\mu_2^0\}= 1/\lambda_2$ at optimality when $\lambda_1 \neq \lambda_2$ for the case when $\min\{1/\lambda_1, 1/\lambda_2\}=1/\lambda_2 < 1/\lambda_1$. Therefore, it always holds true that $\min \{\frac{1}{\lambda_1}, \frac{1}{\lambda_2}\}= \min \{\frac{1}{\mu_1^0}, \frac{1}{\mu_2^0}\}$ if $\lambda_1\neq \lambda_2$.

Next we prove that the optimal water-levels must satisfy condition \eqref{e:optpropty2}. It is straightforward to see that $1/\lambda_1= 1/\lambda_2 \leq 1/\lambda^0$. Moreover, according to the constraints \eqref{e:WLcons1} and \eqref{e:WLcons2}, it is not difficult to see that $1/\lambda_1= 1/\lambda_2 \leq \min\{1/\mu_1^0, 1/\mu_2^0, 1/\mu_{\rm ma}^0\}$ when $1/\lambda_1= 1/\lambda_2$ at optimality. Indeed, if $1/\lambda_1= 1/\lambda_2 > 1/\mu_{\rm ma}^0$, then \eqref{e:WLcons2} cannot be satisfied. If $1/\lambda_1= 1/\lambda_2 > \min\{1/\mu_1^0, 1/\mu_2^0\}$, then \eqref{e:WLcons1} cannot be satisfied. Combining the above two facts, we have $1/\lambda_1= 1/\lambda_2 \leq \min\{1/\mu_1^0, 1/\mu_2^0, 1/\mu_{\rm ma}^0, 1/\lambda^0\}$ when $1/\lambda_1= 1/\lambda_2$ at optimality. For the case that $\min\{1/\mu_1^0, 1/\mu_2^0\}\geq 1/\mu_{\rm ma}^0$, the above constraint can be written as $1/\lambda_1= 1/\lambda_2 \leq \min\{1/\mu_{\rm ma}^0, 1/\lambda^0\}$. For this case,  it is straightforward to see that the achieved sum-rate is not maximized if $1/\lambda_1= 1/\lambda_2 < \min\{1/\mu_{\rm ma}^0, 1/\lambda^0\}$. Therefore, the optimal water-levels must satisfy condition \eqref{e:optpropty2} when $\min\{1/\mu_1^0, 1/\mu_2^0\}\geq 1/\mu_{\rm ma}^0$ given that $1/\lambda_1= 1/\lambda_2$. For the case when $\min\{1/\mu_1^0, 1/\mu_2^0\}< 1/\mu_{\rm ma}^0$, it can be seen that $1/\lambda^0 \leq \min\{1/\mu_1^0, 1/\mu_2^0\}$ given that $1/\lambda_1= 1/\lambda_2$ at optimality. Otherwise, it can be shown that either of the following two results must occur. If $1/\lambda^0 > \min\{1/\mu_1^0, 1/\mu_2^0\}$ and $1/\lambda_1= 1/\lambda_2 \leq \min\{1/\mu_1^0, 1/\mu_2^0\}$, then the sum-rate can be increased. If $1/\lambda^0 > \min\{1/\mu_1^0, 1/\mu_2^0\}$ and $1/\lambda_1= 1/\lambda_2 \geq \min\{1/\mu_1^0, 1/\mu_2^0\}$, then the constraint \eqref{e:WLcons1} cannot be satisfied. Therefore, given that $1/\lambda^0 \leq \min\{1/\mu_1^0, 1/\mu_2^0\}$ for the case when $\min\{1/\mu_1^0, 1/\mu_2^0\}< 1/\mu_{\rm ma}^0$ and $1/\lambda_1= 1/\lambda_2$ at optimality, we have $1/\lambda^0 \leq \min\{1/\mu_1^0, 1/\mu_2^0\}< 1/\mu_{\rm ma}^0$. Consequently, the constraint $1/\lambda_1= 1/\lambda_2 \leq \min\{1/\mu_1^0, 1/\mu_2^0, 1/\mu_{\rm ma}^0, 1/\lambda^0\}$ can be rewritten as $1/\lambda_1= 1/\lambda_2\leq 1/\lambda^0 =\min\{1/\mu_{\rm ma}^0, 1/\lambda^0\}$. It is straightforward to see for this case that $1/\lambda_1= 1/\lambda_2< 1/\lambda^0$ does not maximize the sum-rate. Therefore, it can also be concluded that $1/\lambda_1= 1/\lambda_2= 1/\lambda^0 =\min\{1/\mu_{\rm ma}^0, 1/\lambda^0\}$ when $\min\{1/\mu_1^0, 1/\mu_2^0\}< 1/\mu_{\rm ma}^0$. Combining the above two cases of $\min\{1/\mu_1^0, 1/\mu_2^0\}\geq 1/\mu_{\rm ma}^0$ and $\min\{1/\mu_1^0, 1/\mu_2^0\}< 1/\mu_{\rm ma}^0$, it can be seen that the optimal water-levels always satisfy condition \eqref{e:optpropty2} given that $1/\lambda_1= 1/\lambda_2$.

The above two parts complete the proof of Theorem~1. $\hfill\blacksquare$

\subsection{Proof of Theorem~2}\label{s:PTh2}

The necessity of the constraints \eqref{e:WLcons1} and \eqref{e:WLcons2} is straightforward. It can be seen that the power consumption can be reduced without reducing the sum-rate $R^{\mathrm{tw}}(\mathbf{B}, \mathbf{D})$ when these constraints are not satisfied. The necessity of the constraints \eqref{e:optpropty1} and \eqref{e:optpropty2} is proved in Theorem~1 in Section~\ref{s:PTh1}. Therefore, we next prove the sufficiency of the constraints \eqref{e:WLcons1}, \eqref{e:WLcons2}, \eqref{e:optpropty1}, and \eqref{e:optpropty2}.

We use proof by contradiction. Assume that the above constrains are not sufficient to determine the optimal $\{\lambda_1, \lambda_2\}$ with minimum power consumption among all $\{\lambda_1, \lambda_2\}$'s that maximize the sum-rate $R^{\rm tw}(\mathbf{B}, \mathbf{D})$. Then there exists $\{\lambda_1^\dag, \lambda_2^\dag\}$ satisfying \eqref{e:WLcons} and \eqref{e:optpropty1}-\eqref{e:optpropty2} that maximizes the sum-rate and does not minimize the power consumption. Consequently, at least one of $1/\lambda_1^\dag$ and $1/\lambda_2^\dag$ can be reduced without reducing $R^{\rm tw}(\mathbf{B}, \mathbf{D})$. We consider the following two cases. The first case is when $\lambda_1^\dag \neq \lambda_2^\dag$ while the second case is when $\lambda_1^\dag= \lambda_2^\dag$. In the first case, $\{\lambda_1^\dag, \lambda_2^\dag\}$ satisfies \eqref{e:optpropty1} and it is straightforward to see that reducing $\min\{1/\lambda_1^\dag, 1/\lambda_2^\dag\}$ is not optimal according to Lemma~3. Reducing $\max\{1/\lambda_1^\dag, 1/\lambda_2^\dag\}$, on the other hand, necessarily leads to the decrease of $R^{\rm tw}(\mathbf{B}, \mathbf{D})$ given that \eqref{e:WLcons2} is satisfied. Therefore, reducing either of $1/\lambda_1^\dag$ and $1/\lambda_2^\dag$ results in the decrease of the sum-rate, which contradicts the previous assumption. In the second case, $\{\lambda_1^\dag, \lambda_2^\dag\}$ satisfies \eqref{e:optpropty2}. According to Theorem~2, it is necessary that $1/\lambda_1^\dag= 1/\lambda_2^\dag =\min\{1/\mu_{\rm ma}^0, 1/\lambda^0\}$. From Lemma~2, it can be seen that it is not optimal to reduce only one of $1/\lambda_1^\dag$ and $1/\lambda_2^\dag$. Reducing both of $1/\lambda_1^\dag$ and $1/\lambda_2^\dag$, on the other hand, necessarily leads to the decrease of $R^{\rm tw}(\mathbf{B}, \mathbf{D})$ given that \eqref{e:WLcons2} is satisfied. Therefore, it is impossible that there exists $\{\lambda_1^\dag, \lambda_2^\dag\}$ with $\lambda_1^\dag = \lambda_2^\dag$, satisfying \eqref{e:WLcons} and \eqref{e:optpropty2}, that maximizes the sum-rate while the resulting power consumption can be reduced. Combining the above two cases, it can be seen that the power consumption cannot be reduced given that the $\{\lambda_1^\dag, \lambda_2^\dag\}$ maximizes the sum-rate subject to the relay power limit and satisfies \eqref{e:WLcons} and \eqref{e:optpropty1}-\eqref{e:optpropty2}. This contradicts the assumption that the above constrains are not sufficient to determine the optimal $\{\lambda_1, \lambda_2\}$ with minimum power consumption among all $\{\lambda_1, \lambda_2\}$'s that maximize $R^{\rm tw}(\mathbf{B}, \mathbf{D})$. This completes the proof for Theorem~2. \hfill$\blacksquare$

\subsection{Proof of Theorem~3}\label{s:PTh3}
The optimality of the pair $\{\lambda_1, \lambda_2\}$ obtained
using the algorithm in Table~\ref{t:RPA} is proved in three steps:
A) Steps~2-5 of the algorithm in Table~\ref{t:RPA} find
$\{\lambda_1, \lambda_2\}$ that maximizes $R^{\mathrm{bc}}
(\mathbf{B}, \mathbf{D}^0)$ with minimum power consumption subject
to the constraint in \eqref{e:Rl} and the constraint
\eqref{e:WLcons1}. B) The pair $\{\lambda_1, \lambda_2\}$ obtained
from Steps~2-5 of the algorithm in Table~\ref{t:RPA} needs to be
modified to maximize the objective function in \eqref{e:Rl} with
minimum power consumption. 
Step~6 of the algorithm in
Table~\ref{t:RPA} deals with two cases in which $\{\lambda_1,
\lambda_2\}$ obtained from the previous steps can be simply
modified to obtain the optimal pair $\{\lambda_1, \lambda_2\}$.
C) Step~7 of the algorithm in Table~\ref{t:RPA} deals with the
remaining case which is more complicated and finds the corresponding
optimal pair $\{\lambda_1, \lambda_2\}$ in this case. It is not
difficult to see that the constraint in \eqref{e:Rl} is always
satisfied in any step of the proposed algorithm.
\textcolor{Black}{It can also be
seen that Steps~1, 2 and 6 ensure that \eqref{e:optpropty2} is
satisfied if $\lambda_1=\lambda_2$ at the output of the algorithm
while Steps~3 to 5 ensure that \eqref{e:optpropty1} is satisfied
if $\lambda_1 \neq \lambda_2$ at the output.} Therefore, in the
following we only consider the constraints \eqref{e:WLcons1} and
\eqref{e:WLcons2}, which are equivalent to the constraints in
\eqref{e:RlconsPM}.

A. Steps~2-5 find the pair $\{\lambda_1, \lambda_2\}$ that
maximizes $R(\mathbf{B}, \mathbf{D}^0)$ with minimum power
consumption subject to the constraint \eqref{e:WLcons1}. Note that
the maximum $R(\mathbf{B}, \mathbf{D}^0)$ with minimum power
consumption is achieved by $\hat{R}_{\mathrm{r}1} (\lambda_1) +
\hat{R}_{\mathrm{r}2}(\lambda_2)$ for some specific $\{\lambda_1,
\lambda_2\}$ if \eqref{e:WLcons1} is satisfied. Therefore, it is
equivalent to finding the $\{\lambda_1, \lambda_2\}$ that
maximizes $\hat{R}_{\mathrm{r}1} (\lambda_1) +
\hat{R}_{\mathrm{r}2} (\lambda_2)$ subject to \eqref{e:WLcons1}.
The initial power allocation in Step~1 of the algorithm in
Table~\ref{t:RPA} using $1/\lambda_1=1/\lambda_2=1/\lambda^0$
maximizes $\hat{R}_{\mathrm{r}1}(\lambda_1) +
\hat{R}_{\mathrm{r}2} (\lambda_2)$. Regarding the constraint
\eqref{e:WLcons1},  the following cases are possible.

A-1. $\lambda_i\geq \mu_j^0$, $\forall i$. In this case, the
constraint \eqref{e:WLcons1} is satisfied and $\{\lambda^0,
\lambda^0\}$ is the desired $\{\lambda_1, \lambda_2\}$.

A-2.  $\lambda_i <\mu_j^0$ and $\lambda_j \geq \mu_i^0$. In this
case, the constraint \eqref{e:WLcons1} is not satisfied for $i$.
The relay power consumption can be reduced without decreasing
$R(\mathbf{B}, \mathbf{D}^0)$ by increasing $\lambda_i$ until
$\lambda_i= \mu_j^0$. Then, $R(\mathbf{B}, \mathbf{D}^0)$ can be
increased by decreasing $\lambda_j$ until the relay power limit
is reached or until $\lambda_j=\mu_i^0$.

A-3. $\lambda_i< \mu_j^0, \forall i$. In this case, it is
straightforward to see that the pair $\{\lambda_1, \lambda_2\}$
that maximizes $R(\mathbf{B}, \mathbf{D}^0)$ with minimum power
consumption subject to the constraint \eqref{e:WLcons1} satisfies
$\lambda_i= \mu_j^0, \forall i$.

The above three cases are determined in Step~2. Case A-1 is dealt
with in Step~2 of the algorithm in Table~\ref{t:RPA}. Case A-2 is
dealt with in Steps~3~and~4. Case A-3 is dealt with in
Steps~3~and~5.

B. Steps~6~and~7 of the algorithm in Table~\ref{t:RPA} find the
optimal pair $\{\lambda_1, \lambda_2\}$ that maximizes the
objective function in \eqref{e:Rl} with minimum power consumption.
Since $R^{\rm
ma}(\mathbf{D}^0)<\bar{R}_{1\mathrm{r}}(\mathbf{D}_1^0) +
\bar{R}_{2\mathrm{r}}(\mathbf{D}_2^0)$, it can be seen that
$\lambda_i, \forall i$ should either increase or remain the same
in order to satisfy the constraint \eqref{e:WLcons2} given that
the constraint \eqref{e:WLcons1} is satisfied. Therefore, the
optimal power allocation can be derived by increasing $\lambda_1$
and/or $\lambda_2$, if necessary, based on the power allocation
derived from Steps~1-5. Regarding the constraint
\eqref{e:WLcons2}, the following cases are possible.

B-1. $\lambda_i\geq \mu_\mathrm{ma}^0, \forall i$ or
\bigg($\lambda_i\geq \mu_\mathrm{ma}^0$, $\lambda_j< \mu_\mathrm{ma}^0$
and $\hat{R}_{\mathrm{r}1} (\lambda_1)+\hat{R}_{\mathrm{r}2}
(\lambda_2)\leq R^{\rm ma}(\mathbf{D}^0)$\bigg). In this case, the
constraint \eqref{e:WLcons2} is satisfied and the current
$\{\lambda_1, \lambda_2\}$ is optimal.

B-2. $\lambda_i< \mu_\mathrm{ma}^0, \forall i$ and
$\hat{R}_{\mathrm{r}1}(\lambda_1)+\hat{R}_{\mathrm{r}2}(\lambda_2)
> R^{\rm ma}(\mathbf{D}^0)$. In this case, it is not difficult to
see that it is optimal to simply set $\lambda_i=
\mu_\mathrm{ma}^0, \forall i$.

B-3. $\lambda_i> \mu_\mathrm{ma}^0$, $\lambda_j<
\mu_\mathrm{ma}^0$ and $\hat{R}_{\mathrm{r}1}(\lambda_1) +
\hat{R}_{\mathrm{r}2} (\lambda_2)> R^{\rm ma}(\mathbf{D}^0)$.

Cases B-1 and B-2 are simple and dealt with in Step~6 of the
algorithm in Table~\ref{t:RPA}. It can be shown that in these two cases the constraints \eqref{e:WLcons1} and \eqref{e:WLcons2} are both necessary and sufficient for finding the optimal power allocation in terms of maximizing the sum-rate with minimum power consumption. Case B-3 is dealt with in Step~7.
The optimal strategy in Case B-3, as in Step~7 of the algorithm in
Table~\ref{t:RPA}, is to increase $\lambda_j$ while keeping
$\lambda_i$ unchanged until $\hat{R}_{\mathrm{r}1} (\lambda_1) +
\hat{R}_{\mathrm{r}2}(\lambda_2) = R^{\rm ma}(\mathbf{D}^0)$. In
order to prove that this strategy is optimal, the following three
points are necessary and sufficient.

1. It is optimal to increase $\min\limits_i \{\lambda_i\}$.

2. $\lambda_i=\mu_j^0$ if $\lambda_i > \mu_\mathrm{ma}^0$ and
$\lambda_j< \mu_\mathrm{ma}^0$.

3. At optimality, the increased $\lambda_j$, denoted as
$\lambda_j^{\prime}$, satisfies $\lambda_j < \lambda_j^{\prime} <
\mu_\mathrm{ma}^0$.

The first point states that it is optimal to increase $\lambda_j$
as long as $\lambda_j<\lambda_i$. The second point infers that it
is not optimal to decrease $\lambda_i$. The third point infers
that $\lambda_j^{\prime}$ is always larger than $\lambda_i$ and
therefore it is not optimal to increase $\lambda_i$ at any time.
The first point follows from Lemma~3. For the second point, assume
that $\lambda_i>\mu_j^0$. It follows that $P_{\rm r}^\mathrm{max}$
is used up, i.e., $P_{\rm r}^{\rm max} = \sum\limits_l
\sum\limits_k \big( 1/\lambda_l-1/\alpha_l(k)\big)^+$. Otherwise,
the equality in the constraint \eqref{e:WLcons1} is not achieved
for $i$ and the objective function in \eqref{e:Rl} can be
increased by decreasing $\lambda_i$, which contradicts Steps~1-5
of the algorithm in Table~\ref{t:RPA}. Given that $\lambda_i
> \mu_j^0$ and $P_{\rm r}^\mathrm{max} = \sum\limits_l
\sum\limits_k\big(1/\lambda_l-1/\alpha_l(k)\big)^+$, it can be
proved that $1/\lambda_i\geq1/\lambda_j$. Otherwise, the power
allocation can be proved not optimal based on Lemma~2 because the
objective function in \eqref{e:Rl} is not maximized subject to the
constraint \eqref{e:WLcons1}, which contradicts Steps~1-5 of the
algorithm in Table~\ref{t:RPA}. However, the conclusion that
$1/\lambda_i\geq 1/\lambda_j$ contradicts Case B-3 in which
$\lambda_i> \mu_\mathrm{ma}^0, \lambda_j< \mu_\mathrm{ma}^0$.
Thus, the assumption that $\lambda_i>\mu_j^0$ is invalid. Since
$\lambda_i\geq\mu_j^0$ at the output of Steps~1-5 of the algorithm
in Table~\ref{t:RPA}, we have $\lambda_i=\mu_j^0$. For the third
point, assume that $\lambda_j^{\prime}>\mu_\mathrm{ma}^0$. Then it
follows that $\hat{R}_{\mathrm{r}1} (\lambda_1) +
\hat{R}_{\mathrm{r}2} (\lambda_2)< R^{\rm ma}(\mathbf{D}^0)$ ,
which is not optimal. Therefore, $\lambda_j^{\prime} <
\mu_\mathrm{ma}^0$ at optimality of Case B-3. 

C. Finally, we prove that $\lambda_j^{\prime}$ found in Step~7 of
the algorithm in Table~\ref{t:RPA} for Case B-3 is optimal. The
optimal $\lambda_j^{\prime}$ for Case B-3 is the solution to the
following optimization problem
\begin{subequations} \label{e:eqOPWaterlev}
\begin{align}
&\mathop{\mathbf{min}} \quad \frac{1}{\lambda_j^{\prime}} \\
&\mathbf{\;\;s.t.} \quad\hat{R}_{\mathrm{r}i}(\lambda_i)+\hat{R}_{\mathrm{r}j}
(\lambda_j^{\prime})= R^{\rm ma}(\mathbf{D}^0).
\end{align}
\end{subequations}
Using the definition that
$p_{\mathrm{r}i}(k)=\big({1}/{\lambda_i}-
{1}/{\alpha_i(k)}\big)^{+}$ and
$\mathcal{M}_{\mathrm{r}i}^{+}=\{k|p_{\mathrm{r}i}(k)>0\}$, the
constraint in \eqref{e:eqOPWaterlev} is equal to
\begin{eqnarray}
\hat{R}_{\mathrm{r}i}(\lambda_i)+\sum\limits_{k \in
\mathcal{M}_{\mathrm{r}j}^{+}} \log \frac{\alpha_j(k)}
{\lambda_j^{\prime}} = R^{\rm ma}(\mathbf{D}^0).
\end{eqnarray}
As previously proved, $\lambda_i=\mu_j^0$ in Case B-3, which means
that $\hat{R}_{\mathrm{r}i}(\lambda_{i})=\bar{R}_{j\mathrm{r}}
(\mathbf{D}_{j}^0)$. Thus, the above equation can be written as
\begin{eqnarray}
\sum\limits_{k\in\mathcal{M}_{\mathrm{r}j}^{+}} \log
\frac{\alpha_j(k)}{\lambda_j^{\prime}}= R^{\rm ma}
(\mathbf{D}^0)-\bar{R}_{j\mathrm{r}}(\mathbf{D}_j^0).
\end{eqnarray}
Therefore, the optimal $\lambda_j^{\prime}$ satisfies
\begin{eqnarray}
|\mathcal{M}_{\mathrm{r}j}^{+}|\log\lambda_j^{\prime} = \!
\sum\limits_{k\in\mathcal{M}_{\mathrm{r}j}^{+}} \log\alpha_j(k)-
R^{\rm ma}(\mathbf{D}^0)+\bar{R}_{j\mathrm{r}}(\mathbf{D}_j^0)
\end{eqnarray}
and the optimality of the water level $\lambda_j^{\prime}$ found
in Step~7 of the algorithm in Table~\ref{t:RPA} is proved.

The proof of Theorem~3 is thereby complete. \hfill $\blacksquare $

\end{document}